\documentclass[11pt]{article}
\usepackage{cite}
\usepackage{amsmath,amssymb,amsfonts}
\usepackage{algorithmic}
\usepackage{graphicx}
\usepackage{textcomp}
\usepackage{cite}
\usepackage[a4paper, total={6.5in, 9in}]{geometry}
\usepackage{authblk}
\date{}

\title{A Comparison of CP-OFDM, PCC-OFDM and UFMC for 5G Uplink Communications}

\author[1]{Gayathri Kongara}
\author[2]{Lei Yang}
\author[1]{Cuiwei He}
\author[1]{Jean Armstrong}
\affil[1]{Department of Electrical and Computer Systems Engineering, Monash University, Melbourne, Vic., Australia}
\affil[2]{Faculty of Information Technology, Monash University, Melbourne, Vic., Australia}
\affil[ ]{\textit {\{gayathri.kongara,lei.yang,cuiwei.he,jean.armstrong\}@monash.edu}}

\begin{document}

\maketitle

\begin{abstract}
Polynomial-cancellation-coded orthogonal frequency division multiplexing (PCC-OFDM) is a form of OFDM that has waveforms which are very well localized in both the time and frequency domains and so it is ideally suited for use in the 5G network. This paper analyzes the performance of PCC-OFDM in the uplink of a multiuser system using orthogonal frequency division multiple access (OFDMA) and compares it with conventional cyclic prefix OFDM (CP-OFDM), and universal filtered multicarrier (UFMC).  PCC-OFDM is shown to be much less sensitive than either CP-OFDM or UFMC to time and frequency offsets. For a given constellation size, PCC-OFDM in additive white Gaussian noise (AWGN) requires 3dB lower signal-to-noise ratio (SNR) for a given bit-error-rate, and the SNR advantage of PCC-OFDM increases rapidly when there are timing and/or frequency offsets. For PCC-OFDM no frequency guard band is required between different OFDMA users. PCC-OFDM is completely compatible with CP-OFDM and adds negligible complexity and latency, as it uses a simple mapping of data onto pairs of subcarriers at the transmitter, and a simple weighting-and-adding of pairs of subcarriers at the receiver. The weighting and adding step, which has been omitted in some of the literature, is shown to contribute substantially to the SNR advantage of PCC-OFDM. A disadvantage of PCC-OFDM (without overlapping) is the potential reduction in spectral efficiency because subcarriers are modulated in pairs, but this reduction is more than regained because no guard band or cyclic prefix is required and because, for a given channel, larger constellations can be used.
\end{abstract}


\section{Introduction}
\label{sec:introduction}
The design of the 5G mobile network presents many new challenges not faced by earlier generations of mobile access technology. This is because of the wide range of diverse services 5G will have to support \cite{shafi20175g}. There is the predicted exponential increase in demand for very high bandwidth connection resulting from video streaming and other high data rate applications. At the same time, the emergence of the internet of things (IOT) will produce a very large number of low speed users.  Reconciling these very different types of communication is very challenging and is currently the topic of extensive research \cite{banelli2014modulation, ankarali2017flexible, bodinier2017spectral, ijaz2016enabling, medjahdi2017road, michailow2014generalized, fettweis2009gfdm, premnath2013beyond, schaich2014waveform, farhang2016ofdm}.

Orthogonal frequency division multiplexing (OFDM) in the form of CP-OFDM (cyclic prefix OFDM) has been the basis of many recent wireless communication systems.  The many advantages of CP-OFDM include robustness in the presence of multipath transmission, relative insensitivity to timing offsets, compatibility with multiple-input multiple-output (MIMO) systems and the ability to support multiple access in the form of orthogonal frequency division multiple access (OFDMA). The well-known disadvantages of OFDM include high out-of-band (OOB) power, sensitivity to frequency offset, and high peak-to-average power ratio (PAPR) \cite{armstrong2009ofdm}.

In the current generation of mobile systems, workable solutions have been found which overcome the disadvantages of CP-OFDM. The OOB power can be reduced by leaving some band-edge subcarriers unused and by windowing within the cyclic prefix. Sophisticated synchronization algorithms overcome the problem of frequency sensitivity. A multitude of solutions to the PAPR problem have been described in the literature \cite{han2005overview}. In the LTE uplink a modified form of OFDM called DFT-spread OFDM has been developed to enhance capacity and cell-edge user throughput performance.  The advantages and challenges of using DFT-spread OFDM in 5G are discussed in \cite{priyanto2007initial}. But simple adaptions to these techniques are not adequate for 5G. This has inspired extensive recent research into different waveforms \cite{banelli2014modulation, ankarali2017flexible, bodinier2017spectral, ijaz2016enabling, medjahdi2017road, michailow2014generalized, fettweis2009gfdm, premnath2013beyond, schaich2014waveform, farhang2016ofdm, priyanto2007initial}.

OFDMA  \cite{shafi20175g, farhang2011ofdm} using CP-OFDM may work well in the downlink and meet the demand for higher bandwidth, the requirements of massive MIMO, and the use of higher frequency bands. However, it is not suitable for the uplink of the massive machine-type communication (mMTC) which will result from the IOT. Wunder \textit{et al.} \cite{wunder20145gnow} conclude that the strict synchronization required in OFDMA using CP-OFDM will not be appropriate for the sporadic traffic generated by the IOT for two reasons. Firstly many of these devices will be battery powered and will spend most of the time in a dormant state, awakening only periodically to transmit data, and secondly because of the hardware, time, and energy, that is required for strict synchronization. Hence the search for waveforms that can be used by these devices in the uplink, and which are tolerant to time and frequency offsets. The waveforms which have been considered so far, are well localized in the time and frequency domain but are not necessarily orthogonal. Techniques which have been proposed include filterbank multicarrier (FBMC) \cite{michailow2014generalized, premnath2013beyond, schaich2014waveform, farhang2016ofdm} universal filtered multicarrier UFMC) \cite{vakilian2013universal}, and generalized frequency division multiplexing (GFDM) \cite{michailow2014generalized}. These typically involve some form of frequency domain filtering, or time domain windowing, or some combination of both filtering and windowing.

Polynomial cancellation coded OFDM (PCC-OFDM) is one form of OFDM which is very well localized in both the time and frequency domain, and which is compatible with CP-OFDM and also with MIMO \cite{tang2006mimo, li2010improved}. The first papers on PCC-OFDM were published around two decades ago \cite{armstrong1998polynomial, armstrong1999analysis, zhao1996sensitivity, zhao2001intercarrier} and showed that PCC-OFDM is robust to frequency offset \cite{armstrong1999analysis}. A number of papers quickly followed showing that PCC-OFDM is also robust to time offset \cite{shentu2003effects}, phase noise \cite{ng2005analysis}, and time-varying channels \cite{armstrong1998polynomial, ng2005analysis}, and has a very rapid spectral roll-off \cite{panta2003spectral}. The simplest form of PCC-OFDM involves mapping data to adjacent pairs of subcarriers. This results in substantial cancellation of the frequency domain sidelobes, and a form of windowing in the time domain \footnote{One misconception that has occurred in the past is that PCC-OFDM is simply a form of repetition coding. While repetition coding would give a 3dB gain improvement in an AWGN channel it would not provide the improved performance in the presence of time and frequency offsets that PCC-OFDM demonstrates.}. Its main disadvantage is that it is not spectrally efficient. To overcome this, a form of PCC-OFDM using overlapping symbol periods was developed \cite{armstrong2000performance}.  This uses a concept similar to the weighted-overlap-and-add (WOLA) technique that has recently been proposed for 5G \cite{Qinc2016}. Interest in PCC-OFDM reduced as research in MIMO systems exploded in response to the landmark papers on multiple antenna systems \cite{foschini1998limits,telatar1999capacity}.  However, as we will show the basic form of PCC-OFDM is ideally suited to the uplink in mMTC and has the potential to be a key technology for 5G.

A few papers have recently been published on the application of PCC-OFDM to 5G. Some theoretical analysis and simulation results for PCC-OFDM were presented in \cite{wang2017closed, wang2016waveform}, but these did not include the receiver weighting and adding operation that is required to optimize performance. We will show that the receiver processing gives substantial extra benefit. Experimental work using a software  defined radio testbed has confirmed the potential of PCC-OFDM \cite{kongara2016implementation, kongara2017performance}.

PCC-OFDM maps data onto adjacent pairs of subcarriers.  Data-conjugate ICI cancellation is a closely related system in which data is mapped onto conjugate pairs of subcarriers rather than onto adjacent pairs \cite{ryu2005improved}. It was shown to be less susceptible to phase noise than PCC-OFDM. 

The remainder of this paper is organized as follows. Section \ref{sec:MultiuserOFDMSystems} describes the multiuser systems and the CP-OFDM, PCC-OFDM and UFMC transmitters and receivers that are studied in this paper. In Section \ref{sec:Timeandfrequency}, the time and frequency domain properties of the three waveforms are described, and the intercarrier interference (ICI) caused by time and frequency offsets is analysed. Detailed simulations of the three waveforms in single user systems in the presence of time and frequency offsets are presented in Section \ref{sec:singleuser}. In Section \ref{sec:twousers} the simulations are extended to consider the multiuser case. The many advantages and the possible disadvantages of PCC-OFDM are discussed in Section \ref{sec:advantage}, and finally conclusions are drawn in Section \ref{sec:conclusion}.

\section{Multiuser OFDM Systems}
\label{sec:MultiuserOFDMSystems}
In this section we describe the transmitter and receiver structures for the three forms of OFDM that we analyse and their application in a multiuser system.

\begin{figure}[t] 
\centering
\includegraphics[width=4.0in]{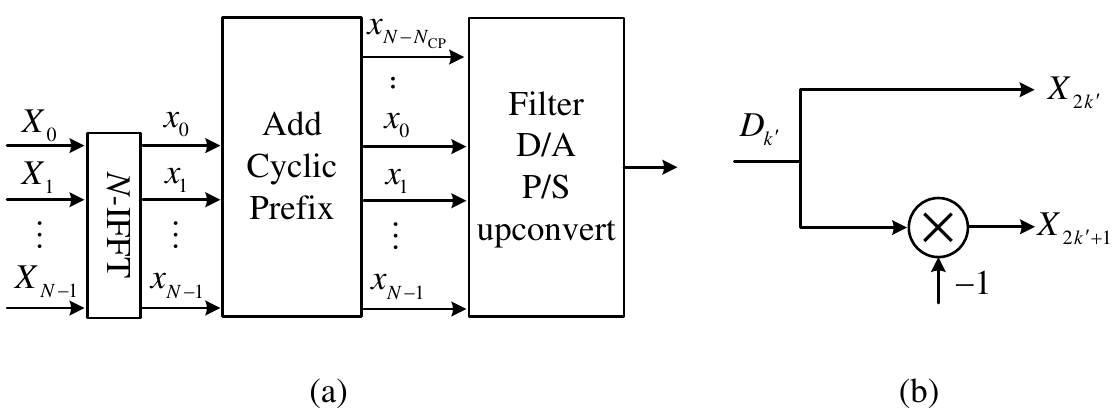}
\caption{OFDM and PCC OFDM transmitter (a) OFDM transmitter and (b) PCC mapping at transmitter.}
\label{fig:fig1}
\end{figure}

\begin{figure}[t] 
\centering
\includegraphics[width=4.0in]{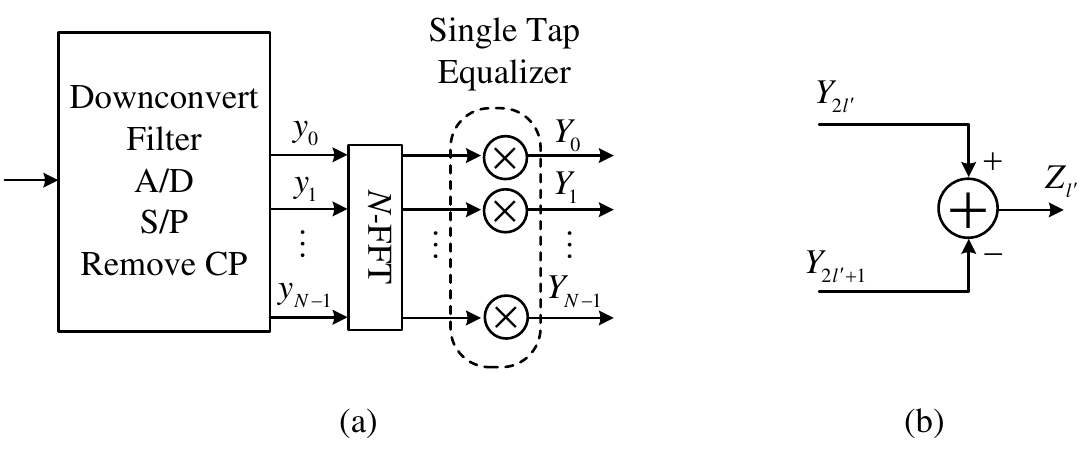}
\caption{OFDM and PCC-OFDM receiver (a) OFDM receiver and (b) PCC OFDM weighting and adding at receiver.}
\label{fig:fig2}
\end{figure}

Fig. 1 shows the key elements of an OFDM transmitter, and the mapping function required for PCC-OFDM. In each symbol period the data to be transmitted is mapped onto a complex vector $\mathbf{X}=\left[ {X}_{0},{X}_{1},{X}_{2}, \cdots, {X}_{N-1} \right]$ which is input to an $N$-point-IFFT. Usually a cyclic prefix is appended to the IFFT output vector before parallel-to-serial conversion, digital-to-analog conversion, filtering and upconversion to a radio frequency carrier.  The corresponding OFDM receiver and PCC weighting and adding block are shown in Fig. 2, where the output from the receiver FFT is the vector $\mathbf{Y}=\left[ {Y}_{0},{Y}_{1},{Y}_{2}, \cdots, {Y}_{N-1} \right]$.

In many applications some subcarriers, for example the band-edge subcarriers, are unused and the corresponding inputs are set to zero. The only difference between a PCC-OFDM transmitter and a simple CP-OFDM transmitter shown is the mapping of data onto adjacent pairs of subcarriers \footnote{To simplify the discussions in the paper we consider the case of mapping onto pairs of subcarriers where lower index is even. This is not necessary.  Any adjacent pairs of subcarriers will provide the same performance gains.}. Each input $D_{{k}'}$ is mapped onto adjacent pairs of subcarriers , so that ${{X}_{2{k}'}}={{D}_{{{k}'}}}$ and ${{X}_{2{k}'+1}}=-{{D}_{{{k}'}}}$ . In a PCC-OFDM receiver, after equalization, the two subcarriers in each pair are combined as shown in Fig. 2(b) to give ${{Z}_{{{l}'}}}={{Y}_{2{l}'}}-{{Y}_{2{l}'+1}}$  . This can be shown to result in a form of receiver windowing and also be equivalent to matched filtering.  Because the noise in the two subcarriers is independent, weighing and adding the two subcarriers in the receiver improves the signal-to-noise ratio (SNR) by 3 dB. We will show that it also further reduces the overall sensitivity of PCC-OFDM to frequency and time offset. Usually no cyclic prefix is required in PCC-OFDM but one can be used if necessary, for example if the same equipment is used for both CP-OFDM and PCC-OFDM. However in most cases the use of a CP will cause slight degradation in the performance of PCC-OFDM.

In this paper, we consider the uplink of a multi-user OFDM based system where multi-user access is achieved by using frequency division multiplexing with different users being allocated different subbands, and where each subband is a group of adjacent OFDM subcarriers. Groups of 12 subcarriers are often considered in the literature and this is what is used in our simulations.  Each user is allocated one or more subbands. To limit interuser interference, guard bands of unused subcarriers may be used between the subcarriers allocated to each user. In this paper we consider two cases: systems with guard bands of 12 subcarriers and systems with no guard band.

\begin{figure}[t] 
\centering
\includegraphics[width=4.7in]{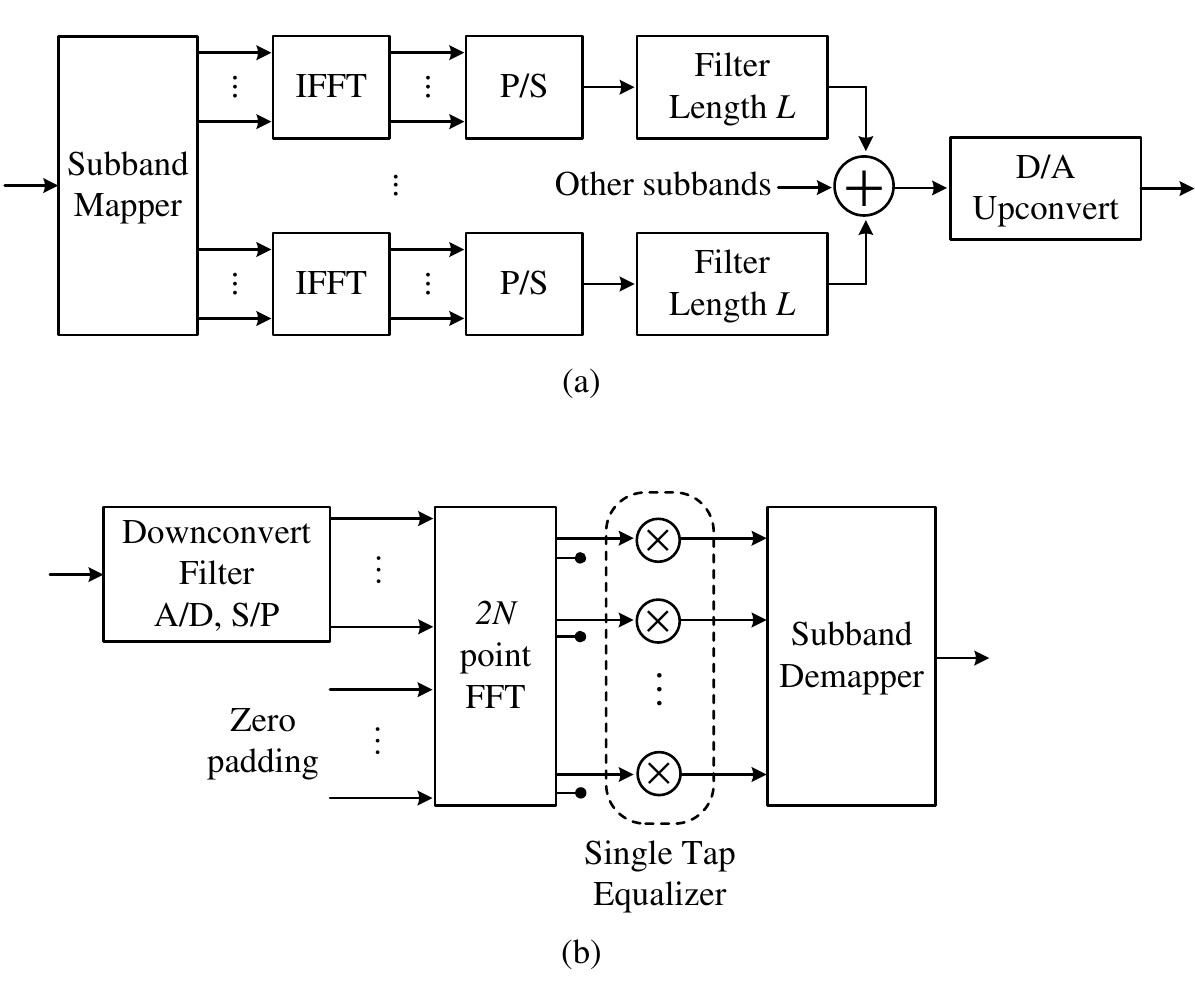}
\caption{UFMC Transmitter and Receiver.}
\label{fig:fig3}
\end{figure}

\begin{figure}[t] 
\centering
\includegraphics[width=5.2in]{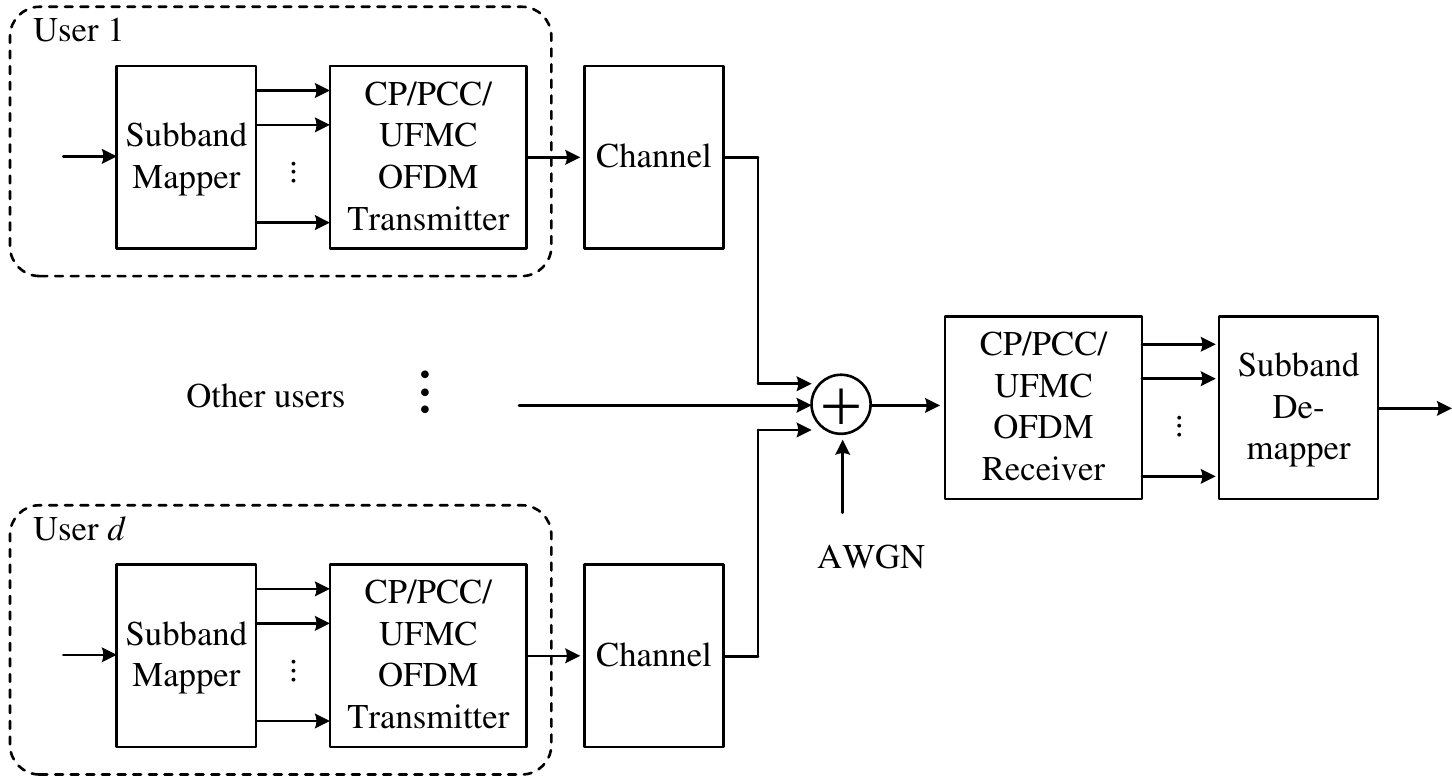}
\caption{Configuration of multi-user uplink.}
\label{fig:fig4}
\end{figure}

The system description for UFMC is more complicated than for CP-OFDM or PCC-OFDM as a separate IFFT and a separate time-domain filter are required for each subband. Fig. 3 shows a UFMC transmitter and receiver.  Each transmitter will in general require multiple $N$-point IFFTs: one IFFT for each subband allocated to that user. To reduce the spectral leakage into other subbands, each subband is separately filtered before the filtered signals representing each subband are combined and transmitted. At the receiver the signal is zero-padded before input to a $2N$ -point FFT. It can be shown that the wanted signals appear on alternate outputs of the IFFT \cite{vakilian2013universal}. These outputs are then equalized before sub-band demapping. A major disadvantage of UFMC is the requirement for multiple FFTs and filters. These increase both the complexity and latency of UFMC systems.

Fig. 4 shows the block diagram that describes the multi-user configuration for the three waveforms that we analyze. A number of different transmitters transmit on different subbands. In this paper we consider the case where all users within a system use the same waveform (i.e. CP-OFDM, PCC-OFDM or UFMC). At each transmitter a subband mapper maps the data onto the subbands allocated to that user. The received signal is the sum of the received signals from each of the users plus additive white Gaussian noise (AWGN). The receiver demodulates the signal and then, based on their allocated subbands, the subband demapper separates the data transmitted by each user. 

\section{Time and frequency domain properties of CP-OFDM, PCC-OFDM and UFMC}
\label{sec:Timeandfrequency}

The relative performances of CP-OFDM, PCC-OFDM and UFMC in a multiuser system depend very much on the time and frequency domain properties of the three waveforms and the level of intercarrier interference (ICI) that any time or frequency offset causes. In this section we describe these properties and the ICI that results from time and frequency domain offsets.

\subsection{Spectra of CP-OFDM, PCC-OFDM and UFMC}
Fig. 5 show the spectra of two 12 subcarrier subbands separated by a guard band 12 subcarriers for systems with IFFT size , $N=256$,  for the three waveforms we consider. Fig. 5(a) show the case of conventional CP-OFDM \footnote{The spectrum for CP-OFDM does not have deep nulls because of the effect of the CP. The extended length of each symbol means that the nulls of different subcarriers occur at slightly different frequencies.}  with CP length , ${N}_{\text{CP}}=32$ . It can be shown that the out-of-band (OOB) spectrum for OFDM falls off with frequency, $f$, as ${1}/{({f}^{2}N)}$ \cite{panta2003spectral}.  The corresponding spectra for PCC-OFDM (with no CP) are shown in Fig. 5(b). The spectral roll-off is much more rapid and it can be shown that the OOB spectrum of PCC-OFDM  fall off as ${1}/{({f}^{4}{N}^{3})}$. This rapid roll-off is an important advantage in subband multiplexing, as it means that there is much less interference between users if the frequencies are not precisely aligned.  Fig. 5(c) shows the spectra for UFMC for a filter length, $L=33$ . Comparing PCC-OFDM and UFMC it can be seen that the OOB power of PCC-OFDM initially falls off more quickly than that of UFMC, but is higher for frequencies further from the subband.

\begin{figure*}[h] 
\centering
\includegraphics[width=6.3in]{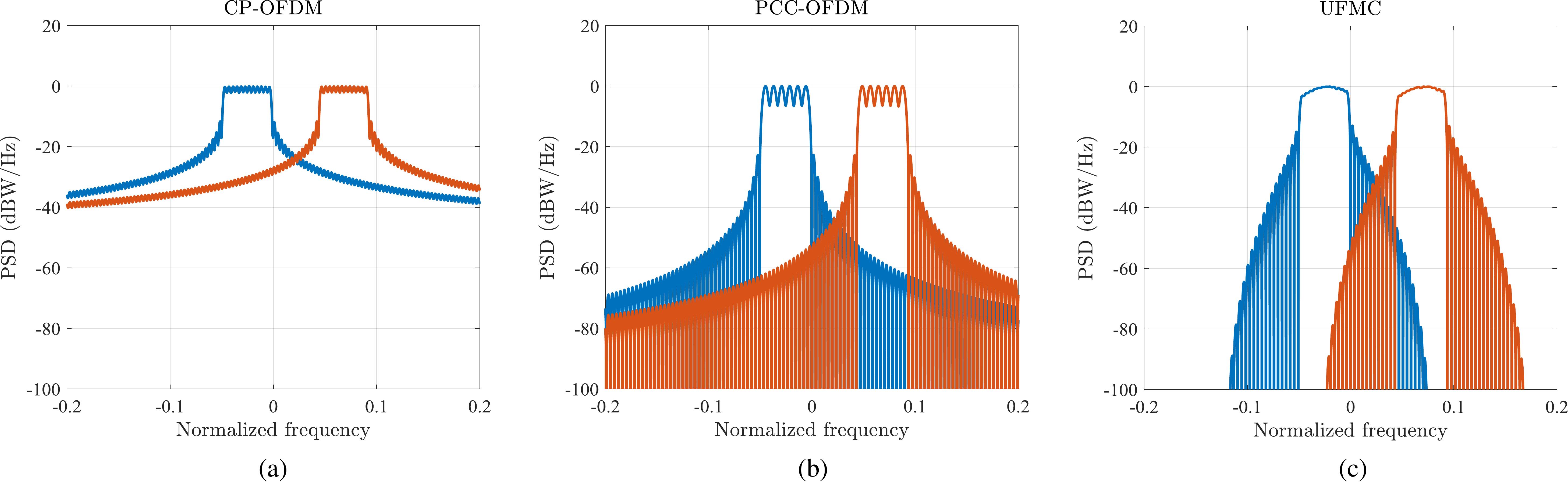}
\caption{Subband spectra for two users for (a) CP-OFDM, (b) PCC-OFDM and (c) UFMC for $N=256$, $N_{\text{CP}}=32$ and $L=33$.}
\label{fig:figure5}
\end{figure*}

It can be seen that for CP-OFDM there is potentially significant interference between subbands if timing or frequency offsets disrupts the strict orthogonality between subcarriers. While there is more overlap for PCC-OFDM than UFMC, the overall performance also depends on the receiver processing. The combining of adjacent subcarriers in PCC-OFDM results in further interference reduction so that, as will be shown in the next section, in the overall system PCC-OFDM outperforms UFMC. The spectrum of UFMC depends on the length of the filter used.  A longer filter will result in more rapid spectral roll-off at the cost of some loss in overall spectral efficiency and increase in signal processing complexity.

\begin{figure}[h] 
\centering
\includegraphics[width=3.5in]{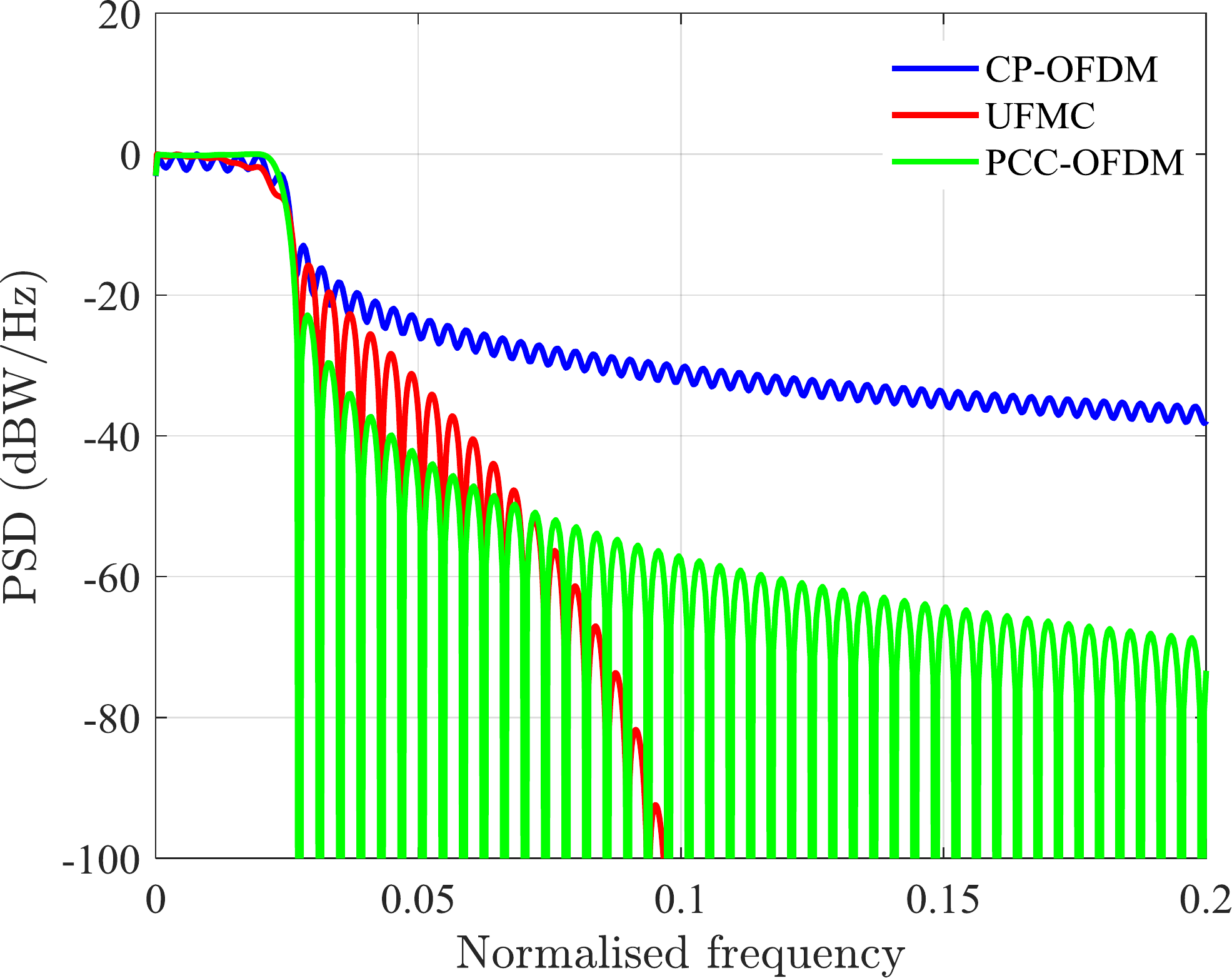}
\caption{Band-edge detail of spectra for CP-OFDM, PCC-OFDM and UFMC, for $N=256$, $N_{\text{CP}}=32$ and $L=33$.}
\label{fig:fig6}
\end{figure}
   
Fig. 6 shows the roll-off of the power spectral density (PSD) at the band edge in more detail.  It shows more clearly the slow roll-off of CP-OFDM, and that PCC-OFDM initially falls off more rapidly, but that UFMC has lower power further from the band edge.  While the spectral roll-off gives some indication of performance, loss of orthogonality in the presence of time and frequency offsets is also an important factor.

\subsection{Time-domain envelopes of CP-OFDM, PCC-OFDM and UFMC}
Fig. 7 shows the time domain envelopes for the three waveforms we consider.  CP-OFDM, shown in Fig.7 (a) has the familiar square window and symbol length of $N+{{N}_{\text{CP}}}=286$ . It can be shown that the PCC-OFDM weighting results in complex windowing given by $\left( 1-\exp \left( {j2\pi l}/{N}\; \right) \right)$ \cite{panta2003spectral}. This has magnitude $\sqrt{2\left( 1-\cos \left( {2\pi l}/{N}\; \right) \right)}$. This windowing effect is clearly shown in Fig. 7(b) which shows the symbol envelope for $N=256$. For PCC the symbol length is equal to the FFT size, as no CP is used. For UFMC the envelope depends on the filter length. Fig. 7(c) shows the result for $N=256$ and  $L=33$ which results in an overall symbol length of $N+L=286$. The filtering in UFMC results in an envelope which tapers at the start and end of the symbol and this reduces the sensitivity of UFMC to time offsets.

\begin{figure}[h] 
\centering
\includegraphics[width=3.5in]{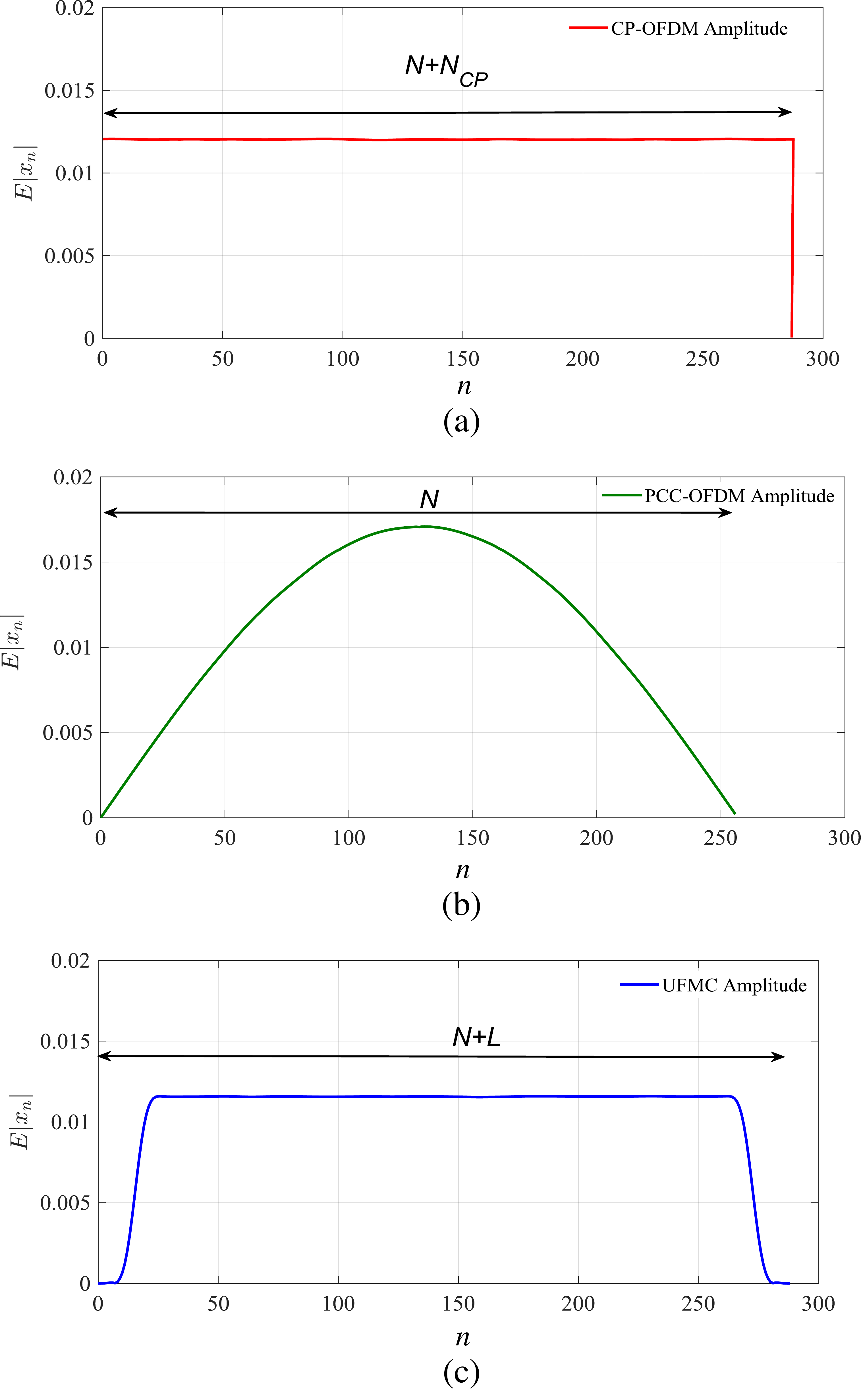}
\caption{Time domain envelopes for  (a) CP-OFDM, (b) PCC-OFDM and (c) UFMC  for $N=256$, ${{N}_{\text{CP}}}=32$ and $L=33$.}
\label{fig:fig7}
\end{figure}

\subsection{Analysis of intercarrier interference caused by time offset}
The degradation in performance of OFDM systems in the presence of time or frequency offsets is a result of the intercarrier interference (ICI) and intersymbol interference (ISI) which these offsets cause. The effect of timing offset on CP-OFDM and PCC-OFDM was analysed in detail in \cite{armstrong2000performance}.  To understand the effects of timing and frequency offset in PCC-OFDM we must introduce some new terminology. In the following, we describe the channel between each PCC-OFDM input $D{}_{{{k}'}}$ and the corresponding output $Z{}_{{{k}'}}$ as a 'subchannel' and the interference between subchannels as intersubchannel interference (ISCHI). Fig. 8 shows three different cases: conventional CP-OFDM, PCC-OFDM without the receiver weighting-and-adding step, and PCC-OFDM with receiver weighting and adding. In each case we do not consider ISI. We consider only the interference within a receiver window by a given transmitted symbol.  Fig. 8(a) shows the ICI in OFDM as a function of time offset ${p}/{N}\;$ for $N=256$  . It shows the power of each output ${{Y}_{l}}$ resulting from an input ${{X}_{k}}=1$ . It can be shown \cite{armstrong2000performance} that for an offset ${p}/{N}\;$
\begin{equation}
Y_{l,k}=\dfrac{1}{N}X_{k}\exp \left( \dfrac{j2\pi kp}{N} \right)\sum_{n=0}^{N-1-p}{\exp \left( \dfrac{j2\pi n(k-l)}{N} \right)}
\label{eq1}
\end{equation}
where ${{Y}_{l,k}}$  is the component of ${{Y}_{l}}$ due to ${{X}_{k}}$. From (1.1) can be seen that the amplitude of ICI in each subcarrier, $\left| {{Y}_{l,k}} \right|$, is a function of only $\left| k-l \right|$ and ${p}/{N}$. For $p=0$ there is no ICI: all of the energy transmitted on a given subcarrier is received on the same subcarrier.  When the time offset increases, the ICI increases and there is substantial ICI even in quite distant subcarriers. Fig. 8(b) shows the results for PCC-OFDM with no receiver weighting. This shows the value of $\left| {{Y}_{l,k}} \right|$ on each received subcarrier when ${{X}_{k}}=1$  and ${{X}_{k+1}}=-1$ are the only non-zero IFFT inputs. Compared with Fig. 8(a) the ICI has substantially reduced. Fig. 8 (c) shows the ISCHI results for PCC-OFDM, when both the transmitter mapping and receiver weighting and adding are considered.  It shows the amplitude of each PCC-OFDM output, $\left| {{Z}_{{l}',{k}'}} \right|$, resulting from an input ${{D}_{k'}}=1$. The ISCHI  is a function only of$\left| {k}'-{l}' \right|$  and ${p}/{N}$.  It can be seen that the level of interference in PCC-OFDM is very much less than that in CP-OFDM. Comparing Fig. 8 (b) and (c) it can also be seen that both the receiver and transmitter PCC-OFDM functions are important in reducing the ISCHI in PCC-OFDM.

\begin{figure*}[h!]
\centering
\includegraphics[scale=0.28]{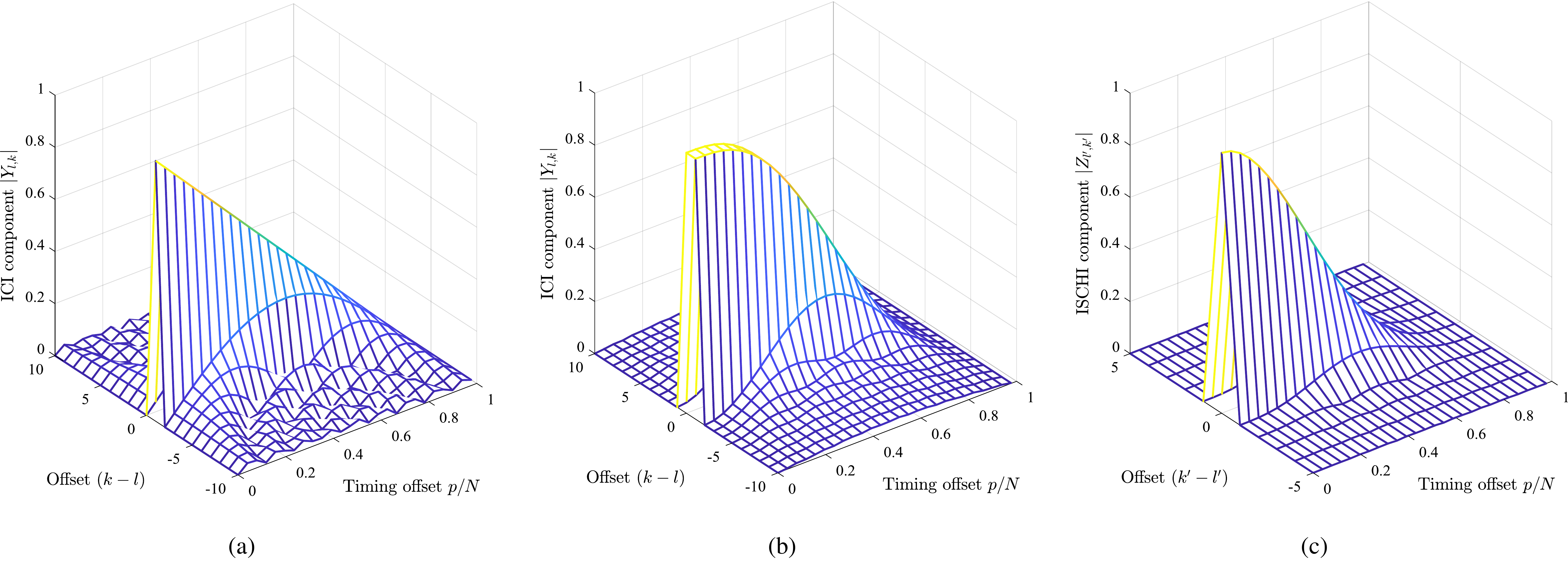}
\caption{ICI and ISCHI as a function of time offset for $N=256$  (a) CP-OFDM, (b) PCC-OFDM no receiver weighting and adding and (c) PCC-OFDM with receiver weighting and adding.}
\label{fig:figure8}
\end{figure*}

\begin{figure*}[h!]
\centering
\includegraphics[scale=0.28]{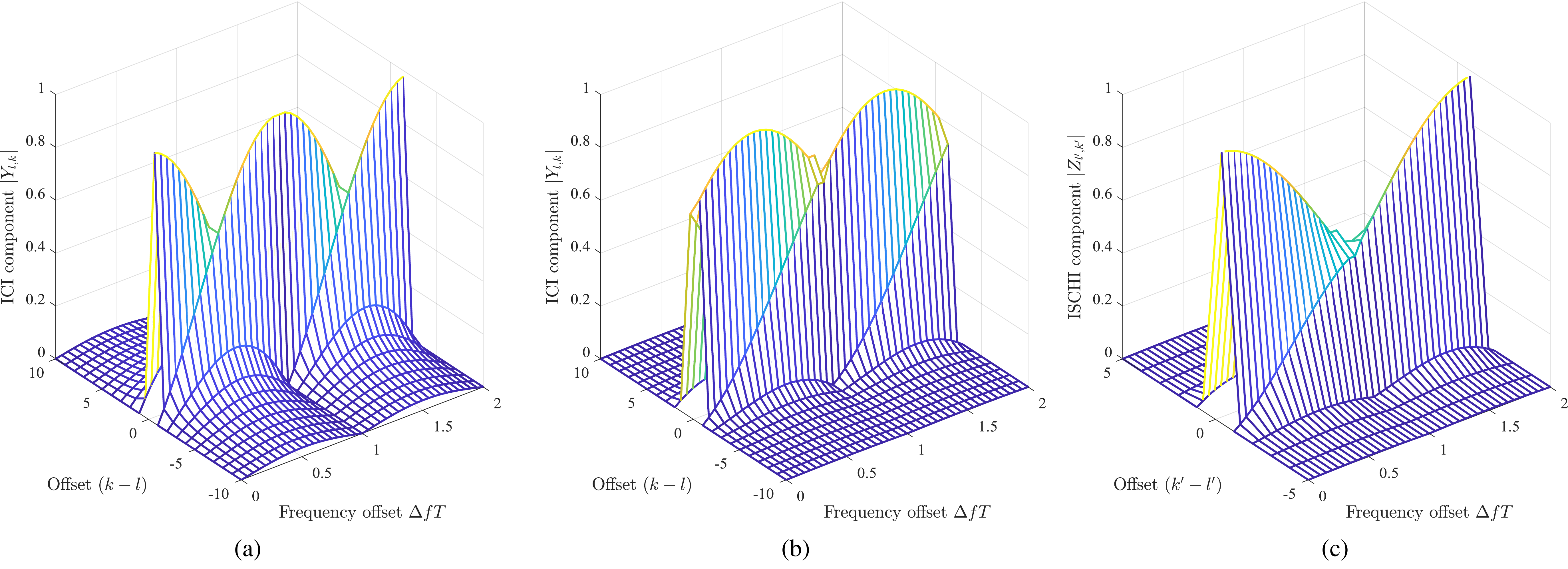}
\caption{ICI and ISCHI as a function of frequency offset for $N=256$ (a) CP-OFDM, (b) PCC-OFDM no receiver weighting and adding and (c) PCC-OFDM with receiver weighting and adding.}
\label{fig:figure9}
\end{figure*}

\subsection{Analysis of intercarrier interference caused by frequency offset}
The ICI due to frequency offset was analysed in detail in \cite{armstrong1999analysis} and it was shown that
\begin{equation}
Y_{l,k}=\dfrac{1}{N}X_{k}\exp \left( j{\theta_{0}} \right)\sum_{n=0}^{N-1}{\exp \left( \dfrac{j2\pi n\left( k-l+\Delta fT \right)}{N} \right)}
\end{equation}	 
where $\Delta f$ is the difference in frequency between the frequency of the receiver local oscillator and the carrier  frequency of the received signal, and ${{\theta }_{0}}$ is the phase offset between the phase of the receiver local oscillator and the carrier phase at the start of the received OFDM symbol. 
Fig. 9 shows the ICI as a function of frequency offset for CP-OFDM, and for PCC-OFDM with and without weighting assuming  ${{\theta }_{0}}=0$ . The normalized frequency offset is varied for $0\le \Delta fT\le 2$ which corresponds to two subcarrier spacings. For CP-OFDM it can be seen in Fig. 9(a) that as $\Delta fT$ varies from 0 to 1, the power gradually shifts from one subcarrier to the next, and that there is significant ICI power across a number of subcarriers. Fig. 9(b) shows the ICI as a function of frequency offset for PCC-OFDM without the receiver weighting and adding step, while Fig. 9(c) shows the ISCHI in PCC-OFDM with receiver weighting-and-adding. It can be seen that as for time offset, both the transmitter mapping and the receiver weighting and adding contribute to the reduction in interference in PCC-OFDM.

\section{Performance of CP-OFDM, PCC-OFDM and UFMC - single-user case}
\label{sec:singleuser}
We now show how the different properties of the three waveforms affect their sensitivity to timing and frequency offsets and to noise in a single-user scenario, and demonstrate the importance of the receiver weighting-and-adding operation in the PCC-OFDM receiver.

\subsection{Performance in an AWGN channel}
We first consider the case of single user in an additive white Gaussian noise (AWGN) channel with no timing or frequency offset between transmitter and receiver. Fig. 10 shows BER results for the three waveforms as a function of ${E_{b}}/{N_{0}}$ where $E_{b}$ is the energy per bit and $N_{0}$ the single-sided noise spectral density. Results are given for 4-QAM, 16-QAM and 64-QAM constellations. For each constellation size, PCC-OFDM requires the lowest ${E_{b}}/{N_{0}}$ of the three waveforms. This is because at the PCC-OFDM receiver the noise in different subcarriers is independent, so combining subcarrier pairs at the receiver reduces the required ${E_{b}}/{N_{0}}$ by 3dB.  The disadvantage of PCC-OFDM is that the mapping of data onto pairs of subcarriers reduces the spectral efficiency, so that for a given constellation size PCC-OFDM carries only slightly more than half the data of CP-OFDM. For CP-OFDM, $E_{b}$ depends on the length of the cyclic prefix. The results are for the case of  $N=256$ and $N_{\text{CP}}=32$. Similarly for FBMC, $E_{b}$ depends on the filter length and the results in Fig. 10 are for $N=256$ and $L=33$. The results for FBMC and CP-OFDM are very similar because $L=N_{\text{CP}}+1$, so the overhead is the same for each system. As expected, for each type of waveform the ${E_{b}}/{N_{0}}$ required for a given BER increases as the constellation size increases.

\begin{figure}[h!]
\centering
\includegraphics[scale=0.37]{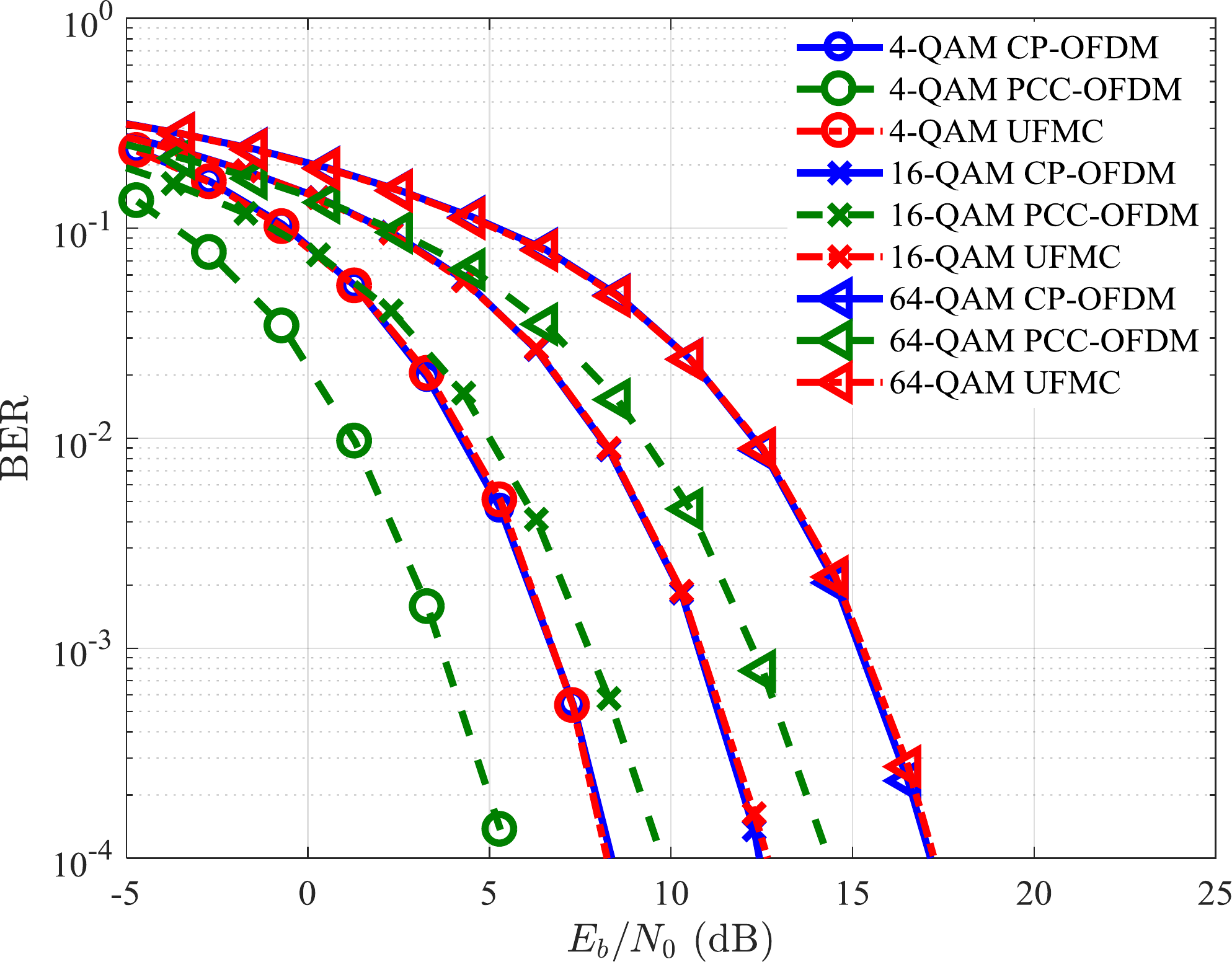}
\caption{BER in an AWGN as a function of ${E_{b}}/{N_{0}}$ for CP-OFDM, PCC-OFDM and UFMC with 4-QAM, 16-QAM and 64-QAM constellations.}
\label{fig:figure10}
\end{figure}

\begin{figure}[h!]
\centering
\includegraphics[scale=0.37]{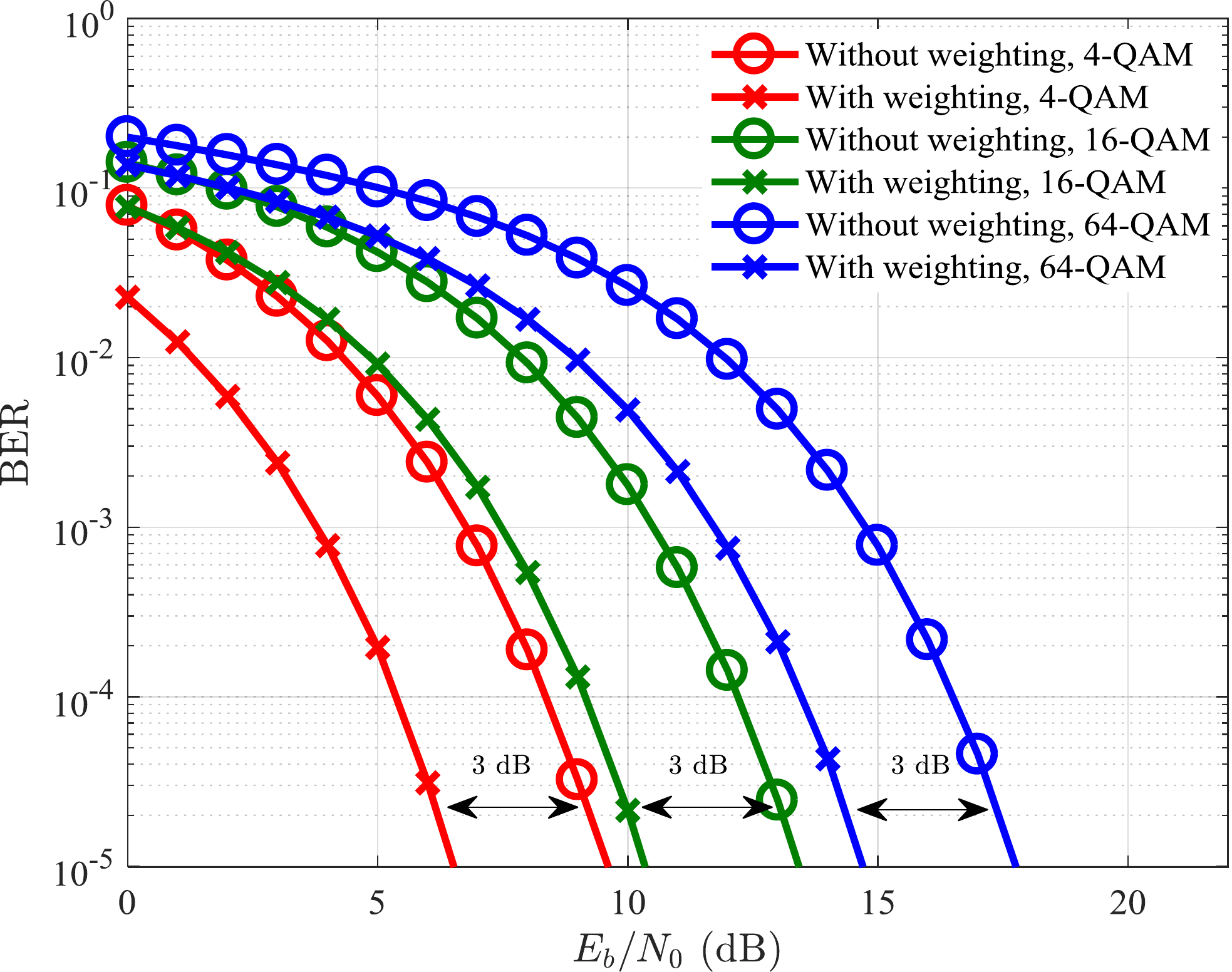}
\caption{BER in an AWGN as a function of ${E_{b}}/{N_{0}}$ for PCC-OFDM with and without weighting and adding in the receiver for  4-QAM, 16-QAM and 64-QAM constellations.}
\label{fig:figure11}
\end{figure}

Fig. 11 demonstrates the importance of the weighting-and-adding operation in a PCC-OFDM receiver. It shows the BER performance for PCC-OFDM without weighting and adding, that is the result if data estimation is based on the received signal on only one of the two PCC subcarriers, and also with weighting and adding of the subcarrier pair.  For each size of constellation it can be seen that the weighting-and-adding operation reduces the required ${E_{b}}/{N_{0}}$ by 3 dB.

\subsection{Performance with timing offset}
We now consider the effect of time offsets between the transmitter and receiver for the single user case.  Fig. 12 shows the BER in AWGN for a receiver offset of $\tau=0.05$ where $\tau$ is measured as  fraction of the OFDM symbol period excluding the CP, and positive $\tau$ represents a delay in the receiver timing relative to the transmitter. As $N=256$ this represents a delay of 13 samples. Comparing Fig. 12 with Fig. 10 it can be seen that this offset does not change the BER for CP-OFDM. This is expected as the offset is less than the CP length of 32 samples.  In contrast the time offset increases the BER for UFMC. This is most clearly seen from the 64-QAM results where UFMC now has a higher BER than CP-OFDM. This is because the receiver window is not aligned with the maximum of the time domain envelope shown in Fig. 7. For PCC-OFDM there is a very slight increase in BER because the first part of each symbol is missed, so not all of the received symbol energy is used to recover the data, but this effect is less than for UFMC as the time domain envelope does not change so quickly. This result is consistent with the effect of timing offset shown in Fig. 8 (c).

\begin{figure}[h!]
\centering
\includegraphics[scale=0.37]{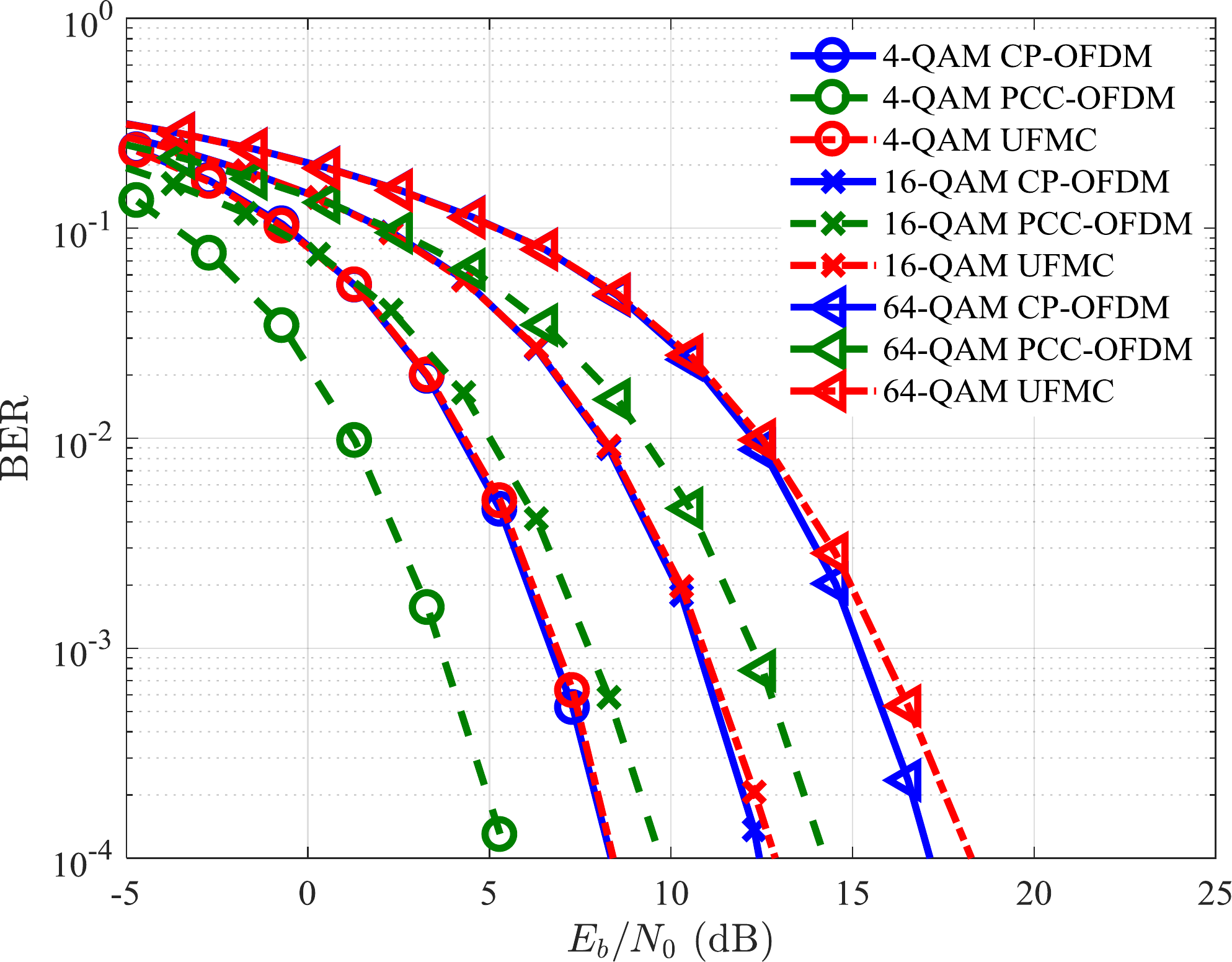}
\caption{BER in an AWGN as a function of  ${E_{b}}/{N_{0}}$ for with time offset $\tau =0.05$ for  CP-OFDM, PCC-OFDM and UFMC with 4-QAM, 16-QAM and 64-QAM constellation.}
\label{fig:figure12}
\end{figure}

\begin{figure}[h!]
\centering
\includegraphics[scale=0.37]{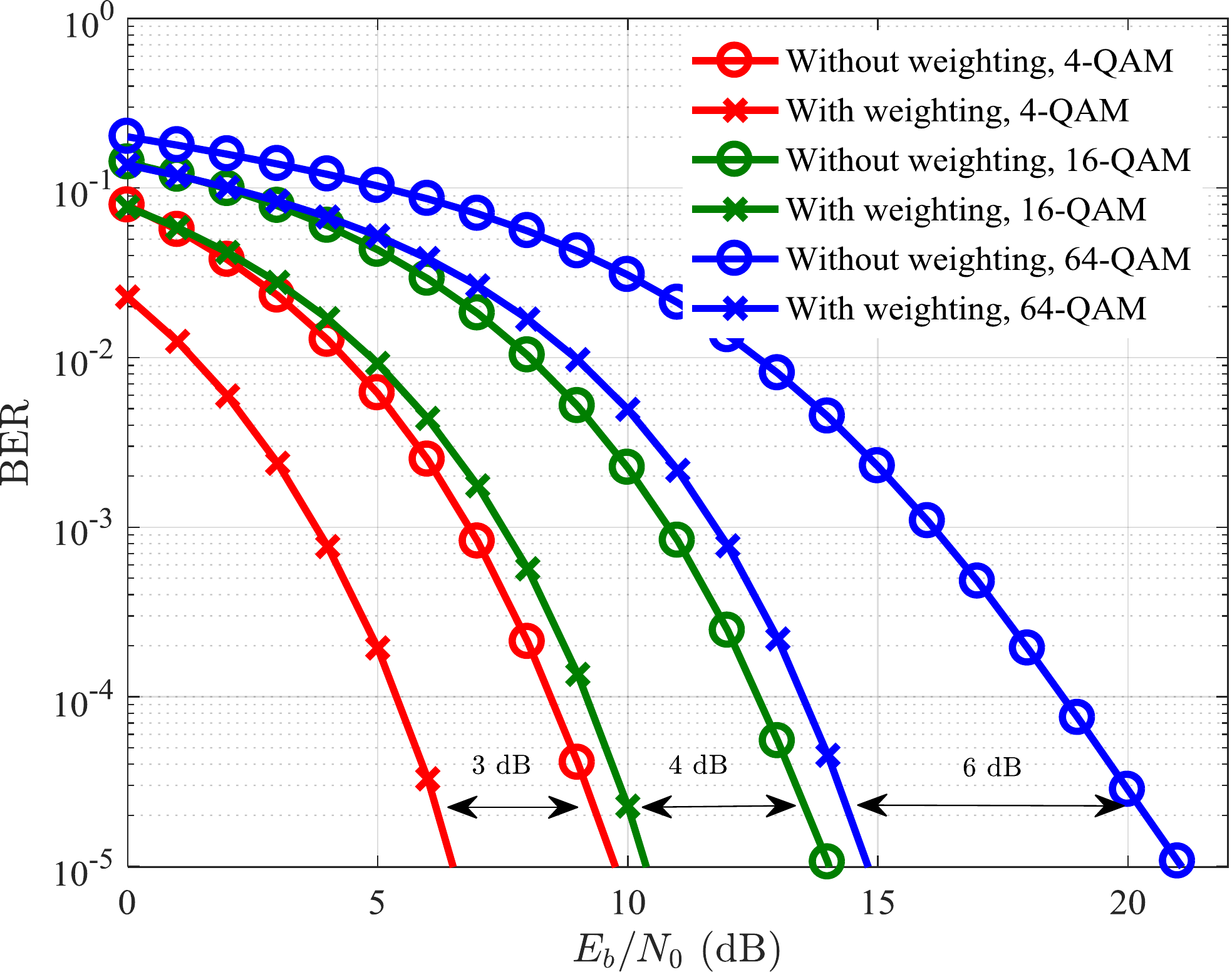}
\caption{BER in an AWGN as a function of ${E_{b}}/{N_{0}}$ for with time offset $\tau =0.05$ for PCC-OFDM with and without weighting and adding in the receiver for  4-QAM, 16-QAM and 64-QAM constellations.}
\label{fig:figure13}
\end{figure}

\begin{figure}[h!]
\centering
\includegraphics[scale=0.37]{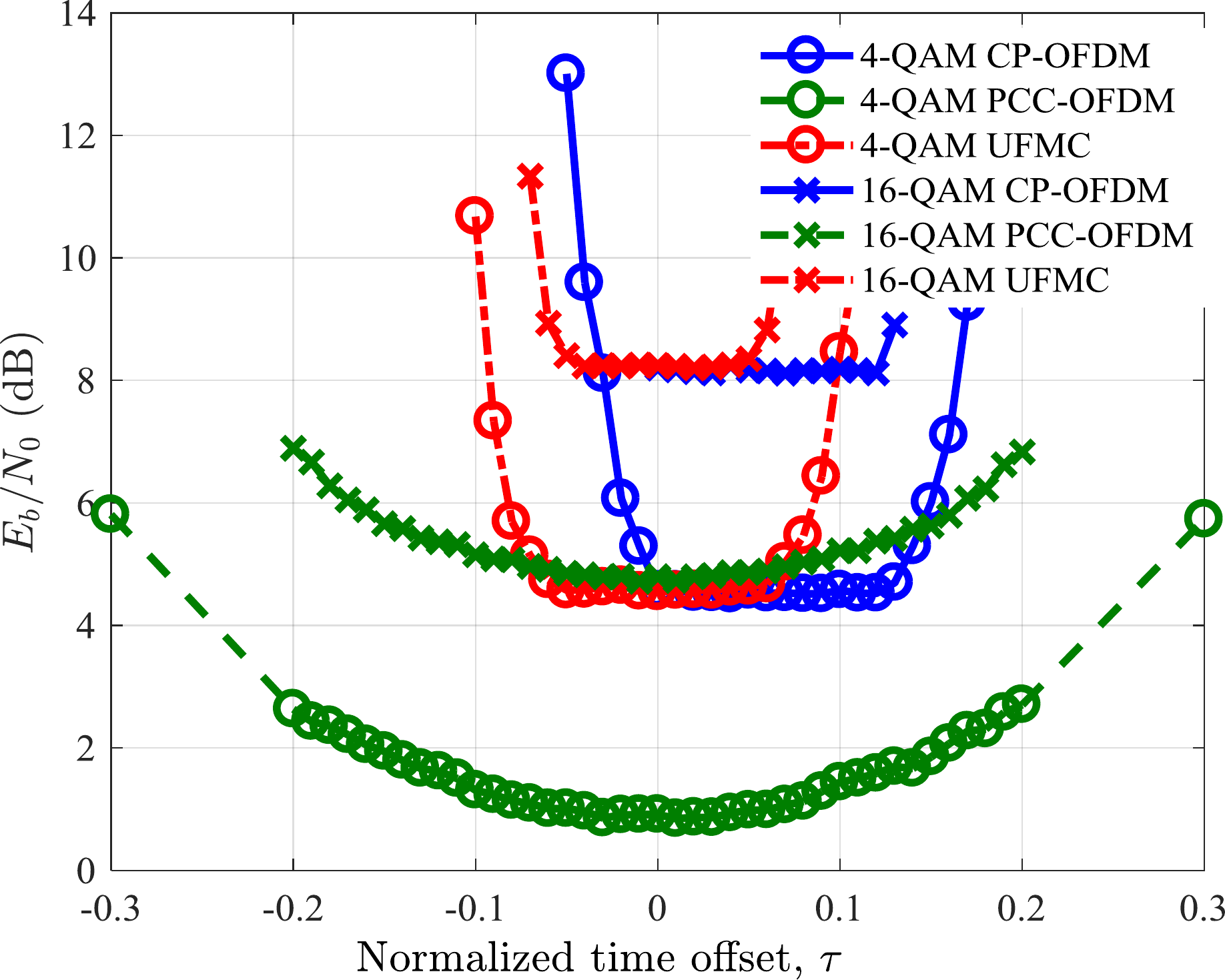}
\caption{${E_{b}}/{N_{0}}$  required for waveforms to achieve a target BER of ${{10}^{-2}}$ as a function of normalized  time offset $\tau $.}
\label{fig:figure14}
\end{figure}

Fig. 13 again shows the importance of the weighting-and-adding operation in the PCC-OFDM receiver.  It shows that it reduces the sensitivity to timing offset as well as providing a 3dB improvement in SNR, it gives an improvement of  up to a 6 dB for 64-QAM with a timing offset of $\tau=0.05$.

We now explore the effect of varying the timing offset. Fig. 14 shows how the performance of each waveform varies as a function of time offset, $\tau$. It compares the required ${E_{b}}/{N_{0}}$ for CP-OFDM, PCC-OFDM and UFMC for a target BER of $10^{-2}$ for 4-QAM and 16-QAM. For CP-OFDM the BER is constant as long as the delay is within the cyclic prefix length, but increases rapidly for $\tau<0$, or $\tau>0.125$. In contrast PCC-OFDM is much less sensitive to time offset and degrades only slowly. Even offsets of $\tau =\pm 0.2$ require an increase of less than 2dB to maintain a BER of $10^{-2}$. The performance of UFMC is relatively constant over the smaller range $-0.08<\tau<0.08$ but degrades rapidly for larger time offsets.

\subsection{Performance with frequency offset - single user case}
Frequency offsets between transmitter and receiver will also degrade the BER performance. Fig. 15 shows the effect of a normalized frequency offset, $\Delta fT=0.05$. Comparing Fig. 15 with the AWGN results shown in Fig. 10, it can be seen that frequency offset has negligible effect on PCC-OFDM for all constellation sizes, while the performance of CP-OFDM and UFMC are significantly degraded, with the BER plateauing for 64-QAM constellations. Fig. 16 shows the effect of varying frequency offsets. Both UFMC and CP-OFDM degrade rapidly as frequency offset increases, while even normalized frequency offsets of 0.2 have little effect on PCC-OFDM.

A point to note is that both for large time offsets and large frequency offsets 16-QAM PCC-OFDM outperforms 4-QAM UFMC, and for small offsets has approximately equal performance.  This means that the loss in spectral efficiency of PCC-OFDM due to the mapping of data onto two subcarriers, can potentially be regained by using bigger constellations.

\begin{figure}[h!]
\centering
\includegraphics[scale=0.37]{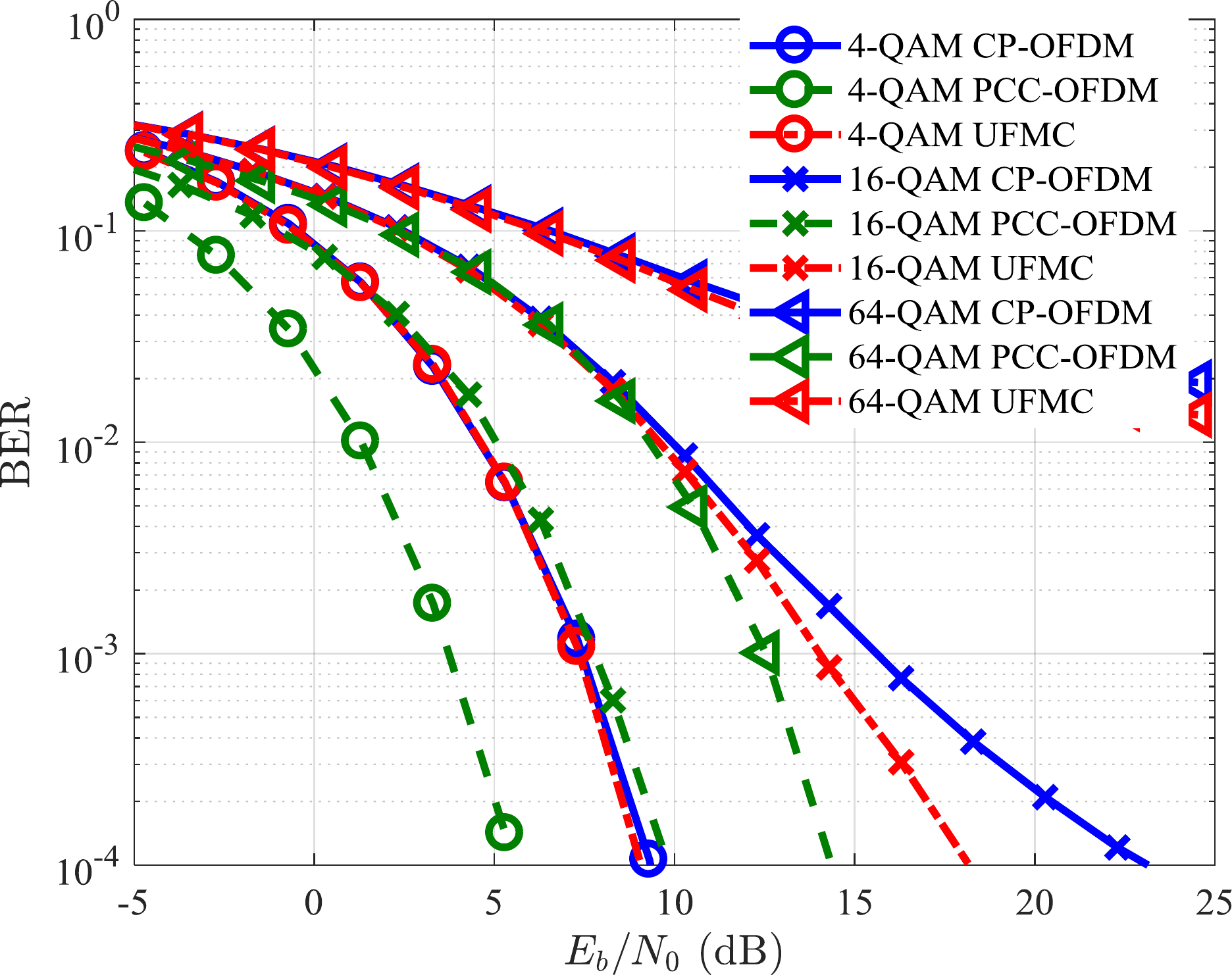}
\caption{BER comparison of three waveforms for varying constellation sizes affected by a normalized frequency offset, $\Delta fT=0.05$.}
\label{fig:figure15}
\end{figure}

\begin{figure}[h!]
\centering
\includegraphics[scale=0.37]{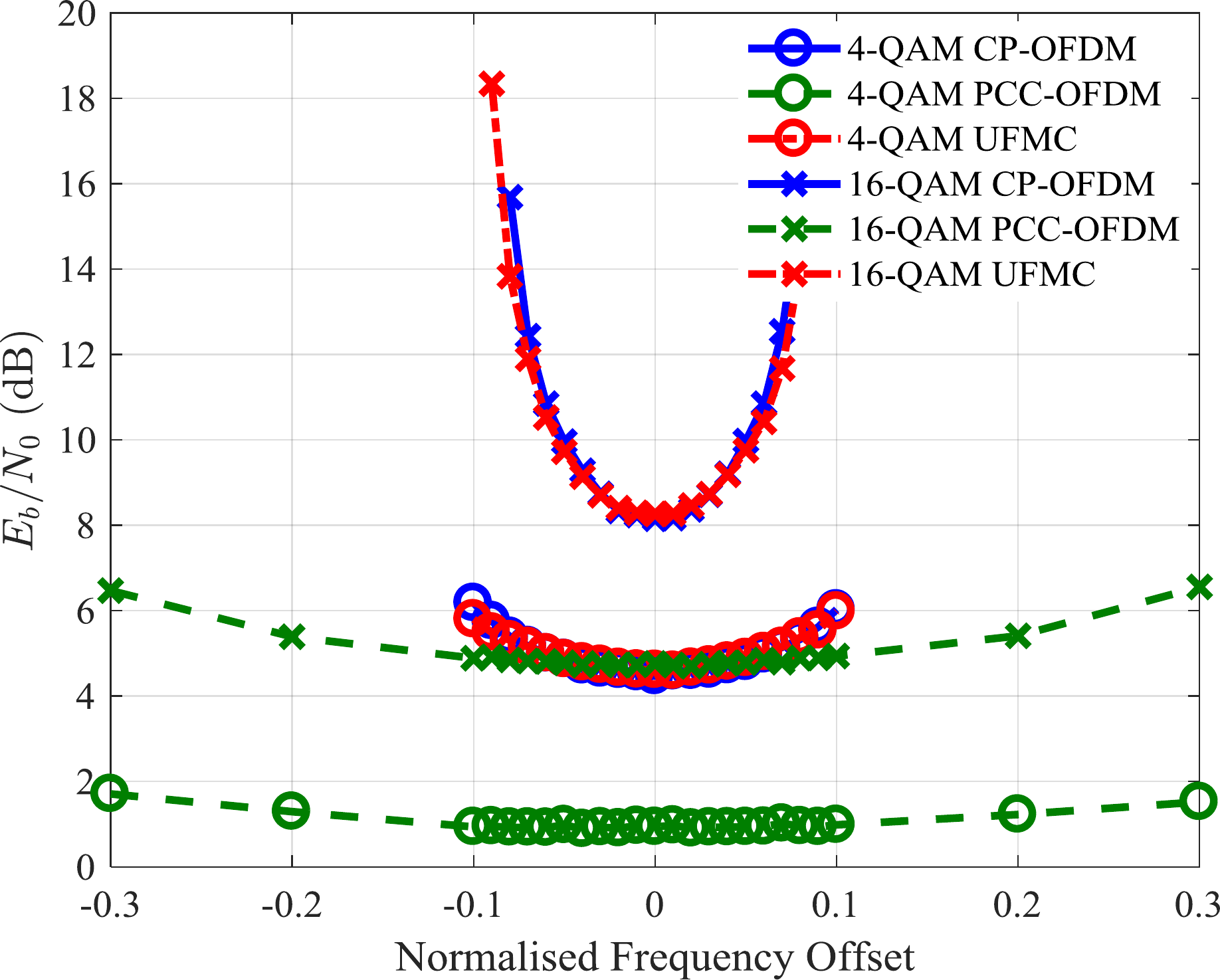}
\caption{${E_{b}}/{N_{0}}$ required for waveforms to achieve a target BER of ${{10}^{-2}}$ as a function of normalized frequency offset, $\Delta fT$.}
\label{fig:figure16}
\end{figure}

\begin{figure}[h!]
\centering
\includegraphics[scale=0.37]{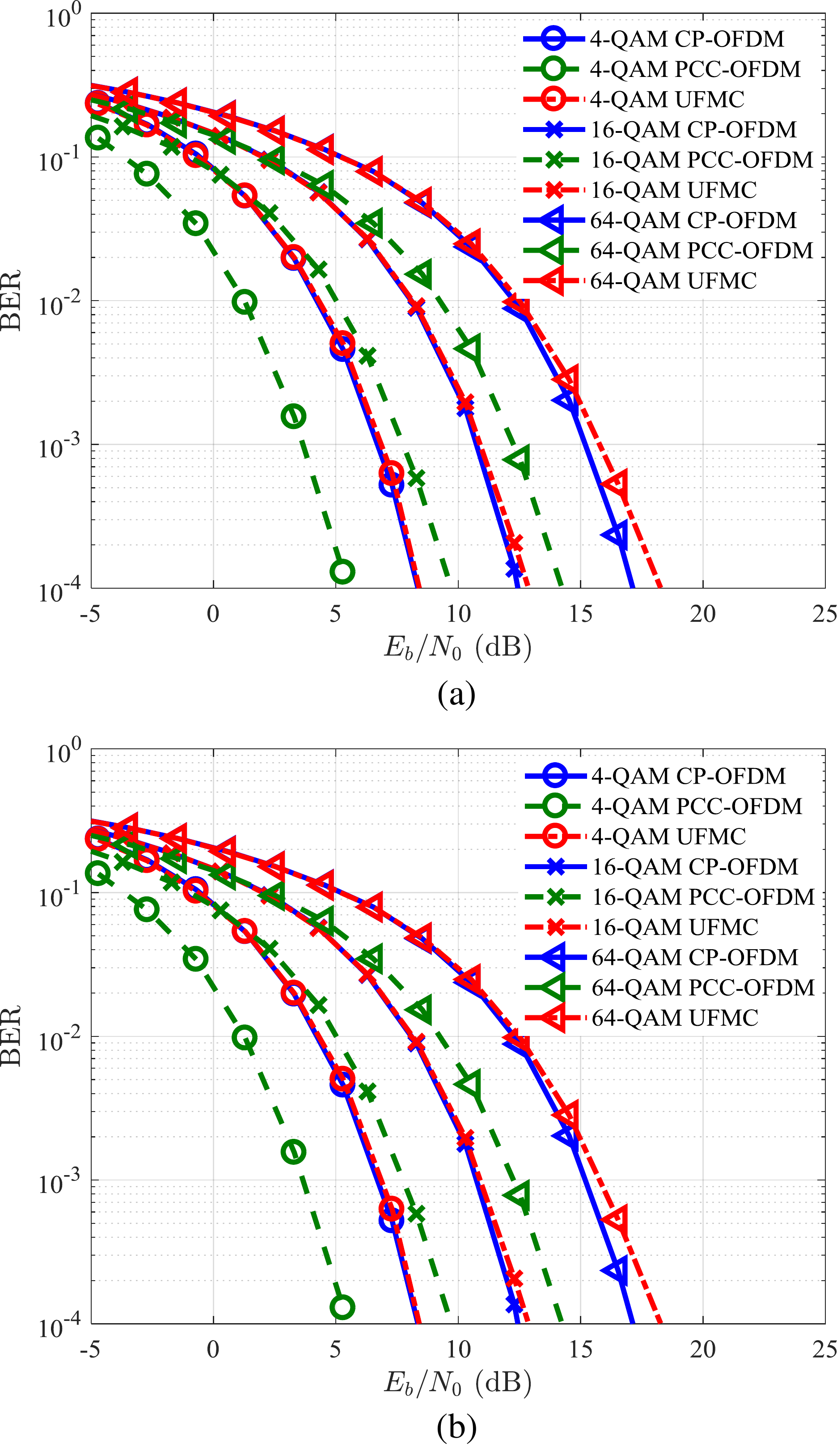}
\caption{Two user BER results for normalized timing offset of User 2, $\tau=0.05$ (a) guard band of 12 subcarriers, (b) no guard band.}
\label{fig:figure17}
\end{figure}

\begin{figure}[h!]
\centering
\includegraphics[scale=0.37]{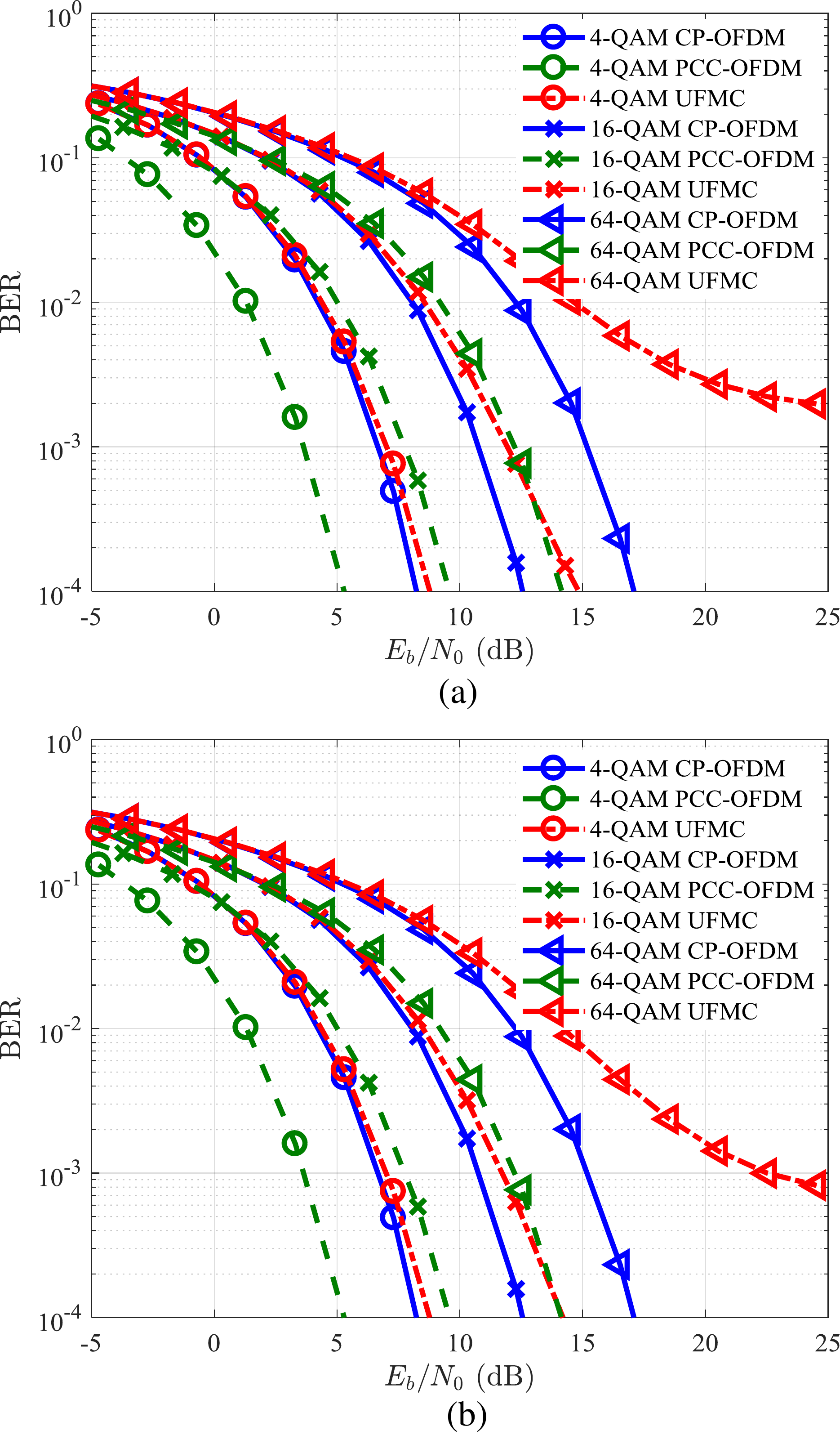}
\caption{Two user BER results for normalized timing offset of User 2, $\tau =0.05$.  User 2 received power 10 dB greater than User 1 (a) guard band of 12 subcarriers, (b) no guard band.}
\label{fig:figure18}
\end{figure}

\section{Performance of CP-OFDM, PCC-OFDM and UFMC - two-user case}
\label{sec:twousers}
As we are ultimately interested in the application of PCC-OFDM for uplink multiple access, we now investigate the sensitivity of the three waveforms to time and frequency offsets in a two-user scenario. A number of configurations are possible, but here we calculate the BER for User 1 for the case where the receiver is synchronized to User 1, so that there is no timing or frequency offset between User 1 and the receiver, but User 2 may have a timing or frequency offset. Whereas in the previous section the ICI as 'same user ICI': the ICI was caused by subcarriers allocated to one user, in this section the impairment may also be due to 'other user ICI', which is the ICI caused by subcarriers allocated to a different user. Similarly for ISI.  As can be seen in Fig. 8, the level of ICI is strongly dependent on the spacing between the subcarriers concerned, so we consider both cases where there is a guard band between the subcarriers allocated to each user and cases where there is no guard band.

\subsection{Performance of two-user system in AWGN channel}

Simulations were performed for two users for 12 subcarrier subbands with a 12 subcarrier guard band and an AWGN channel. The BER results were identical to the single user AWGN case shown in Fig. 10, and this was also the case when there was no guard band. This was the expected result for CP-OFDM and PCC-OFDM as in the absence of time or frequency offset the subcarriers are strictly orthogonal. However there was also no observable degradation for UFMC. In general the powers of the signals received from different users will not be equal, to explore the effect of a strong interfering signal, the received power of User 2 was increased by 10 dB. There was still no observable degradation: the BER plots were the same as Fig. 10.

\subsection{Performance  of two-user system with timing offset}
Fig. 17 shows the BER performance when User 2 has a time offset of $\tau=0.05$. For a 12 subcarrier guard band comparing Fig. 17 (a) with Fig. 10 it can be seen that there is a very slight degradation in the performance of 64-QAM UFMC. Removing the guard band (Fig. 17  (b)) causes no observable increase in degradation of UFMC. Fig. 18 shows the effect of increasing the received power of User 2 by 10 dB.  The key point is that even with no guard band and a higher power interfering signal there is no observable change in the performance of PCC-OFDM.  There is also no change for CP-OFDM as this time offset is within the cyclic prefix. However the performance of 64-QAM UFMC degrades significantly. A slightly surprising result is that for this case the no guard band 64-QAM UFMC has better performance than 64-QAM UFMC with a guard band.  This is because with no synchronization  errors the ICI in UFMC falls on the odd subcarriers but timing offsets disrupt this.

\begin{figure}[h!]
\centering
\includegraphics[scale=0.37]{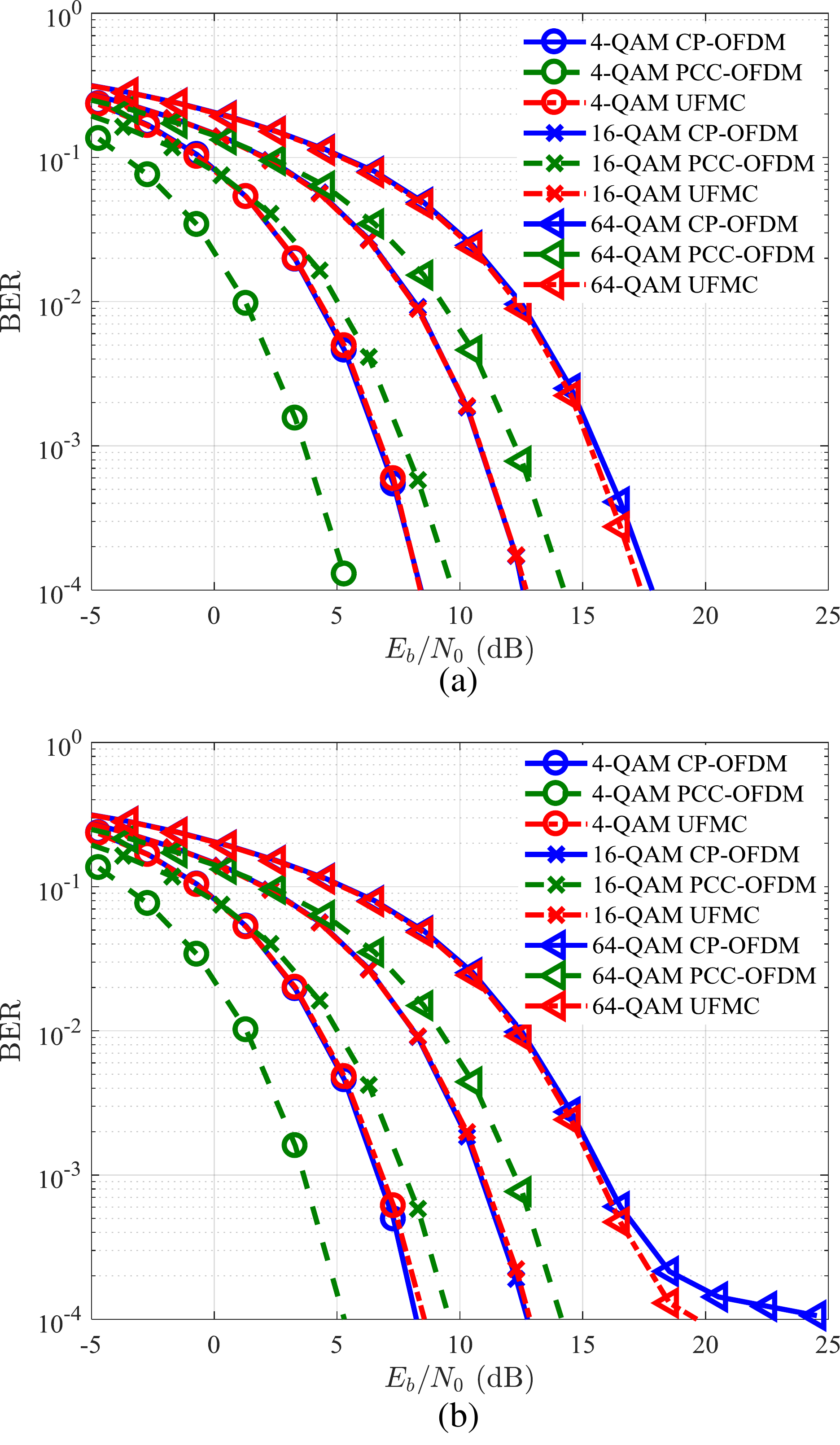}
\caption{Two user BER results for normalized frequency offset of User 2, $\Delta fT=0.05$ (a) guard band of 12 subcarriers, (b) no guard band.}
\label{fig:figure19}
\end{figure}

\begin{figure}[h!]
\centering
\includegraphics[scale=0.37]{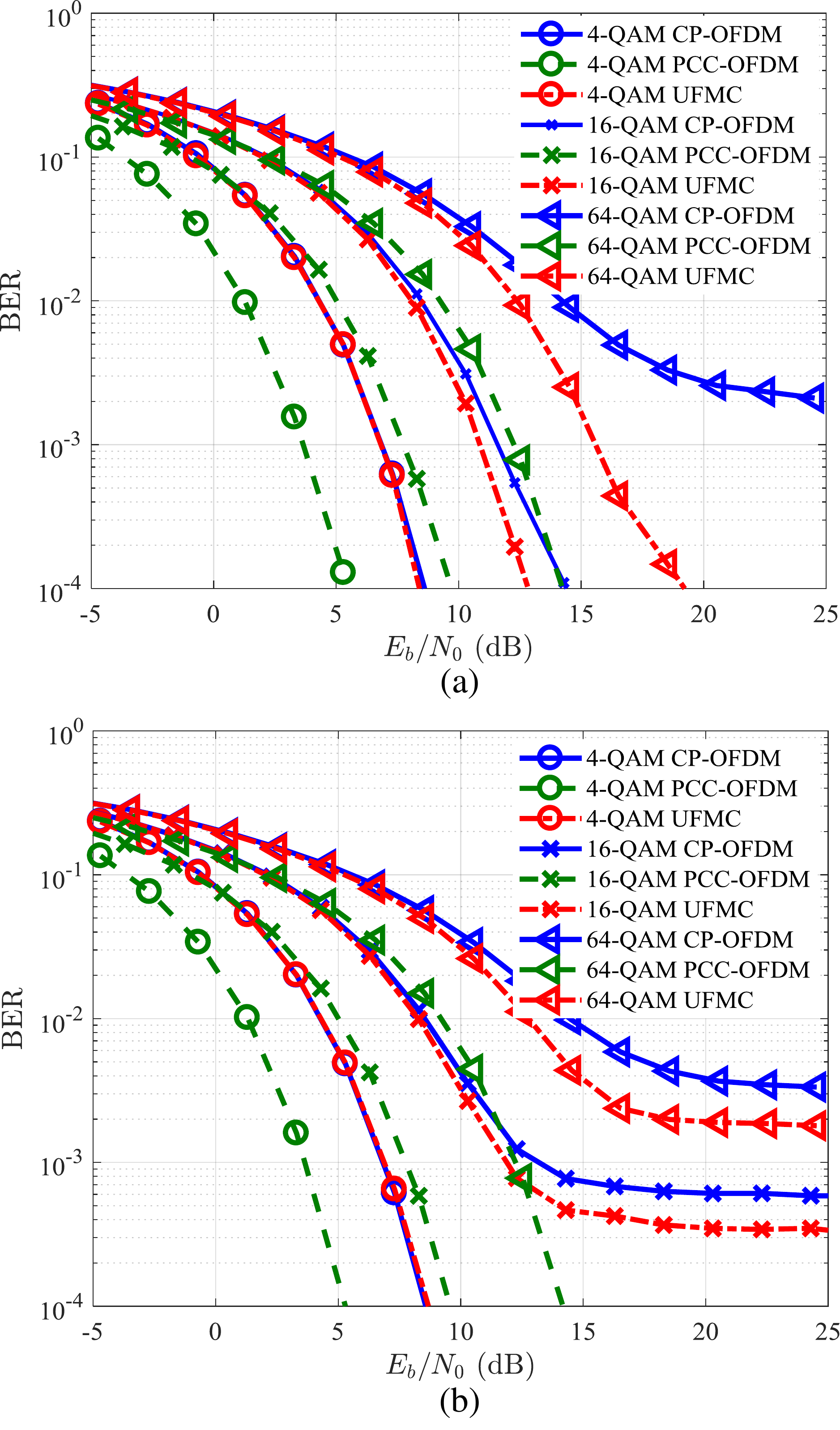}
\caption{Two user BER results for normalized frequency offset of User 2, $\Delta fT=0.05$, User 2 received power 10 dB greater than User 1 (a)	guard band of 12 subcarriers, (b) no guard band.}
\label{fig:figure20}
\end{figure}

\subsection{Performance  of two-user system with frequency offset} 
Fig. 19 shows the BER performance when User 2 has a frequency offset of $\Delta fT=0.05$. For a 12 subcarrier guard band comparing Fig. 19 (a) with Fig. 10 it can be seen that there is a very slight degradation in the performance of 64-QAM CP-OFDM. With no guard band (Fig. 19 (b)) the performance of 64-QAM CP-OFDM degrades further and 64-QAM UFMC also shows some change. Fig. 20 shows the effect of increasing the received power of User 2 by 10 dB.  The key point is that even with no guard band and a higher power interfering signal there is no observable change in the performance of PCC-OFDM.  Fig. 21 shows the effect of increasing the frequency offset. Now even with a guard band (Fig. 21 (a))  the frequency offset degrades the performance of CP-OFDM and UFMC.  Fig. 22 shows that only for large frequency offset and for a high power interfering signal does PCC-OFDM show any degradation, and this is only observable for the largest constellation.

\begin{figure}[h!]
\centering
\includegraphics[scale=0.37]{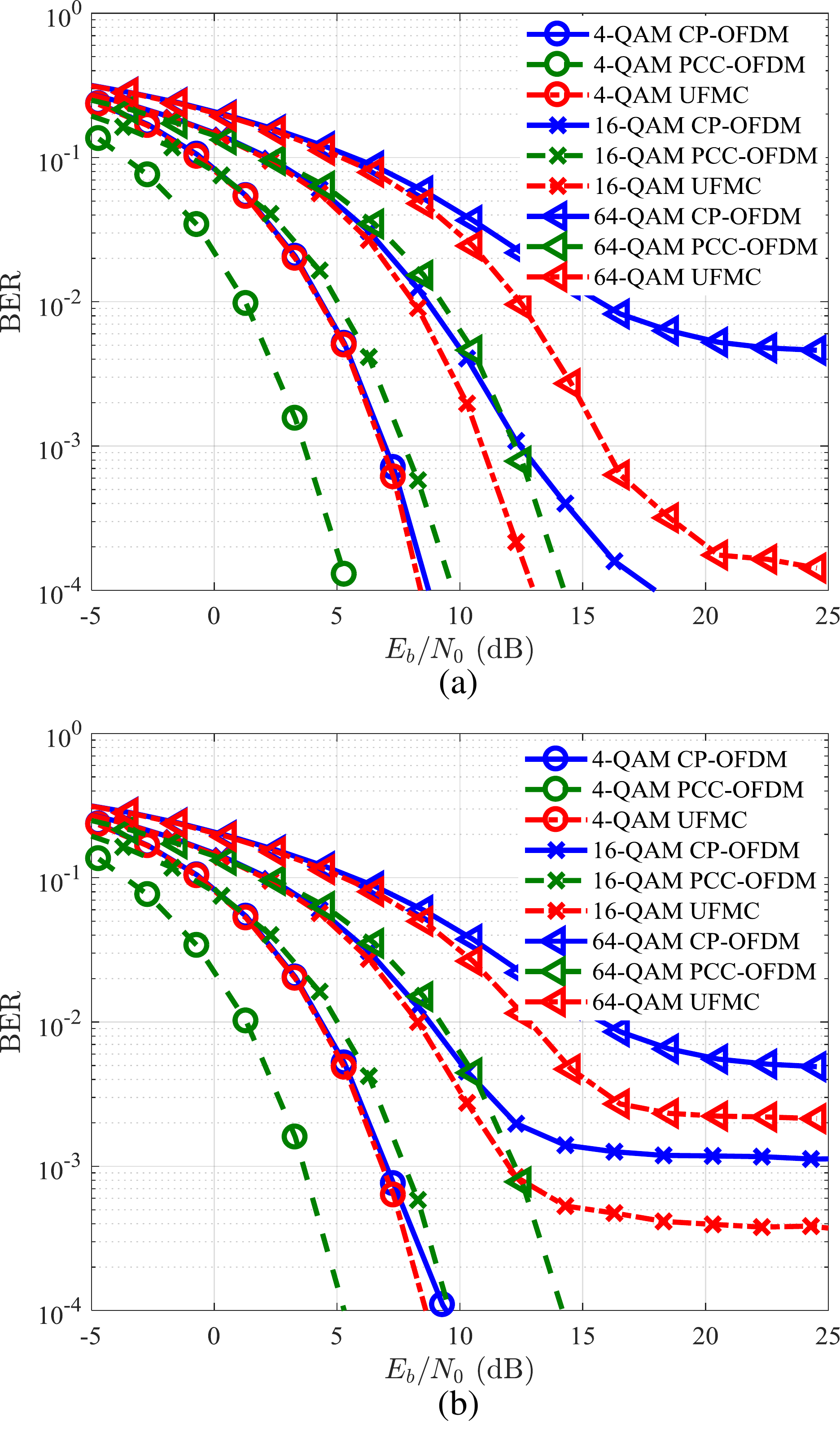}
\caption{Two user BER results for normalized frequency offset of User 2, $\Delta fT=0.2$ (a)	guard band of 12 subcarriers, (b) no guard band.}
\label{fig:figure21}
\end{figure}

\begin{figure}[h!]
\centering
\includegraphics[scale=0.37]{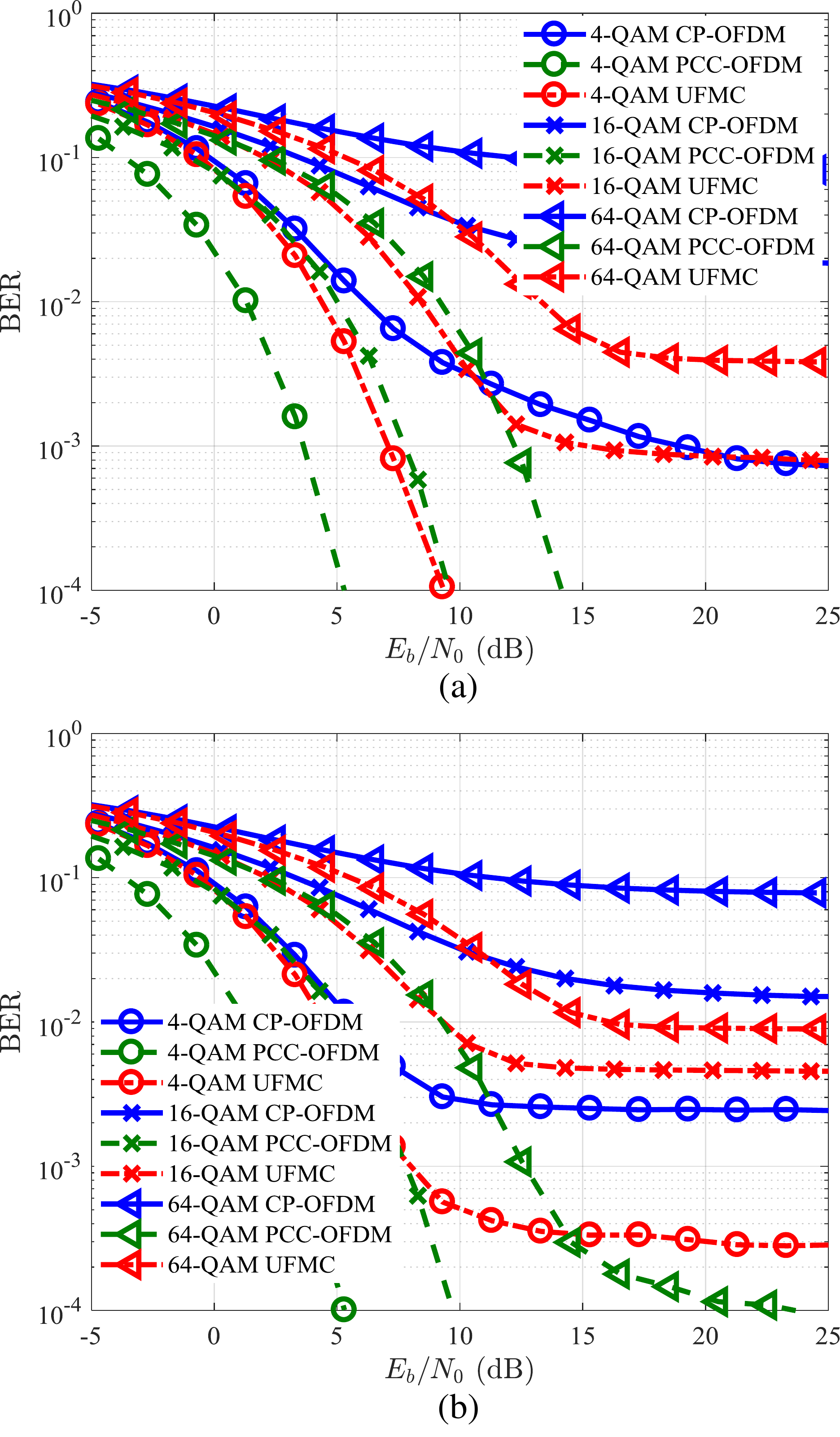}
\caption{Two user BER results for normalized frequency offset of User 2, $\Delta fT=0.2$, User 2 received power 10 dB greater than User 1
(a)	guard band of 12 subcarriers, (b) no guard band.}
\label{fig:figure22}
\end{figure}

\subsection{Performance  of two-user system with time and frequency offset}
Finally we investigate the performance when User 2 has both time and frequency offset. Fig. 23 that all three waveforms perform well for small time and frequency offsets even with no guard band. Comparing Fig. 23 which shows the results for $\Delta fT=0.05$, $\tau=0.05$ with Fig. 10 it can be seen that only 64-QAM CP-OFDM and 64-QAM UFMC show any change. Similarly, Fig. 24 shows that even for $\Delta fT=0.2$, $\tau=0.05$ and with no guard band PCC-OFDM shows no degradation. 

\begin{figure}[h!]
\centering
\includegraphics[scale=0.37]{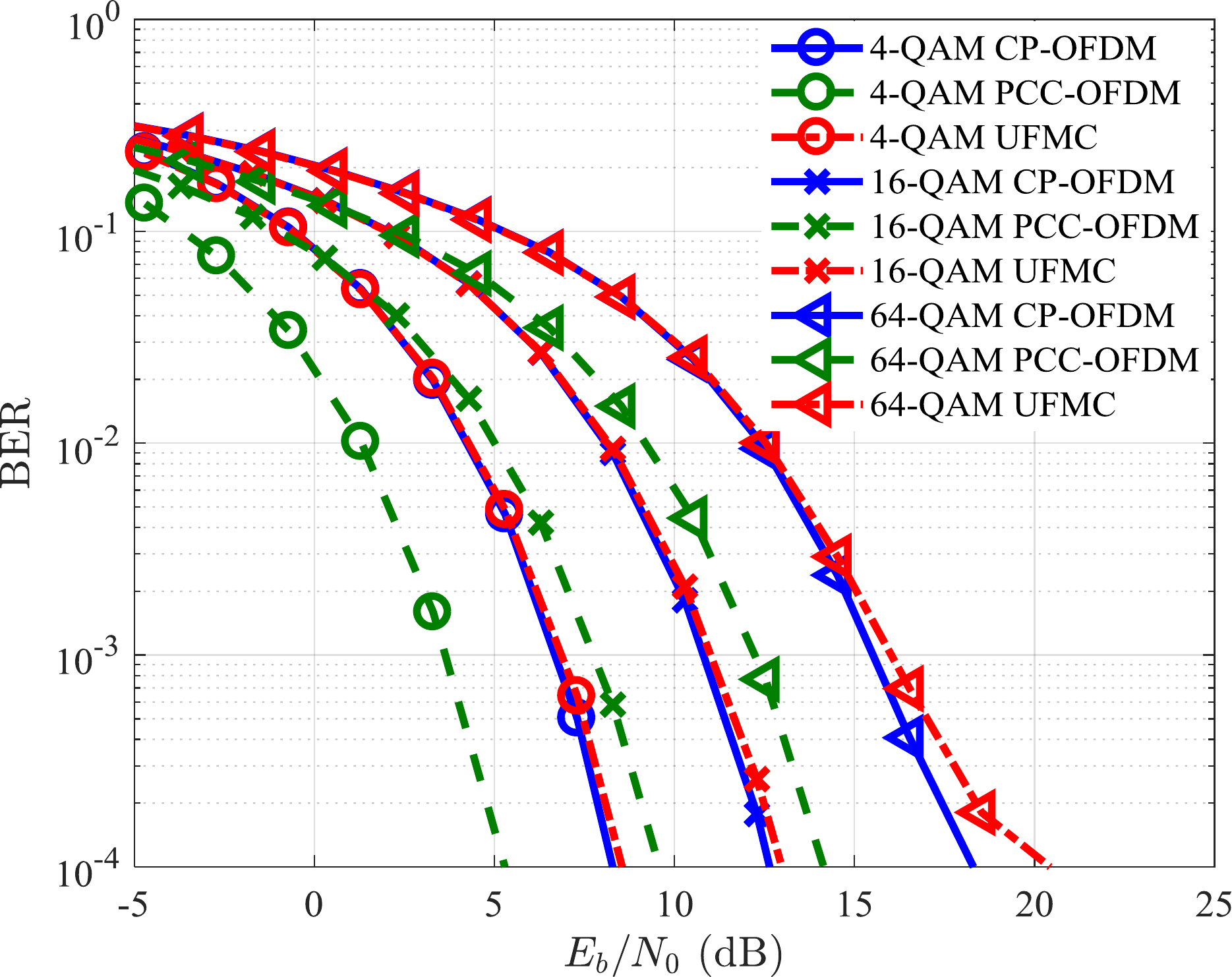}
\caption{Two user BER results for normalized frequency offset of User 2, $\Delta fT=0.05$, and normalize time offset of User 2, $\tau =0.05$. No guard band.}
\label{fig:figure23}
\end{figure}

\begin{figure}[h!]
\centering
\includegraphics[scale=0.37]{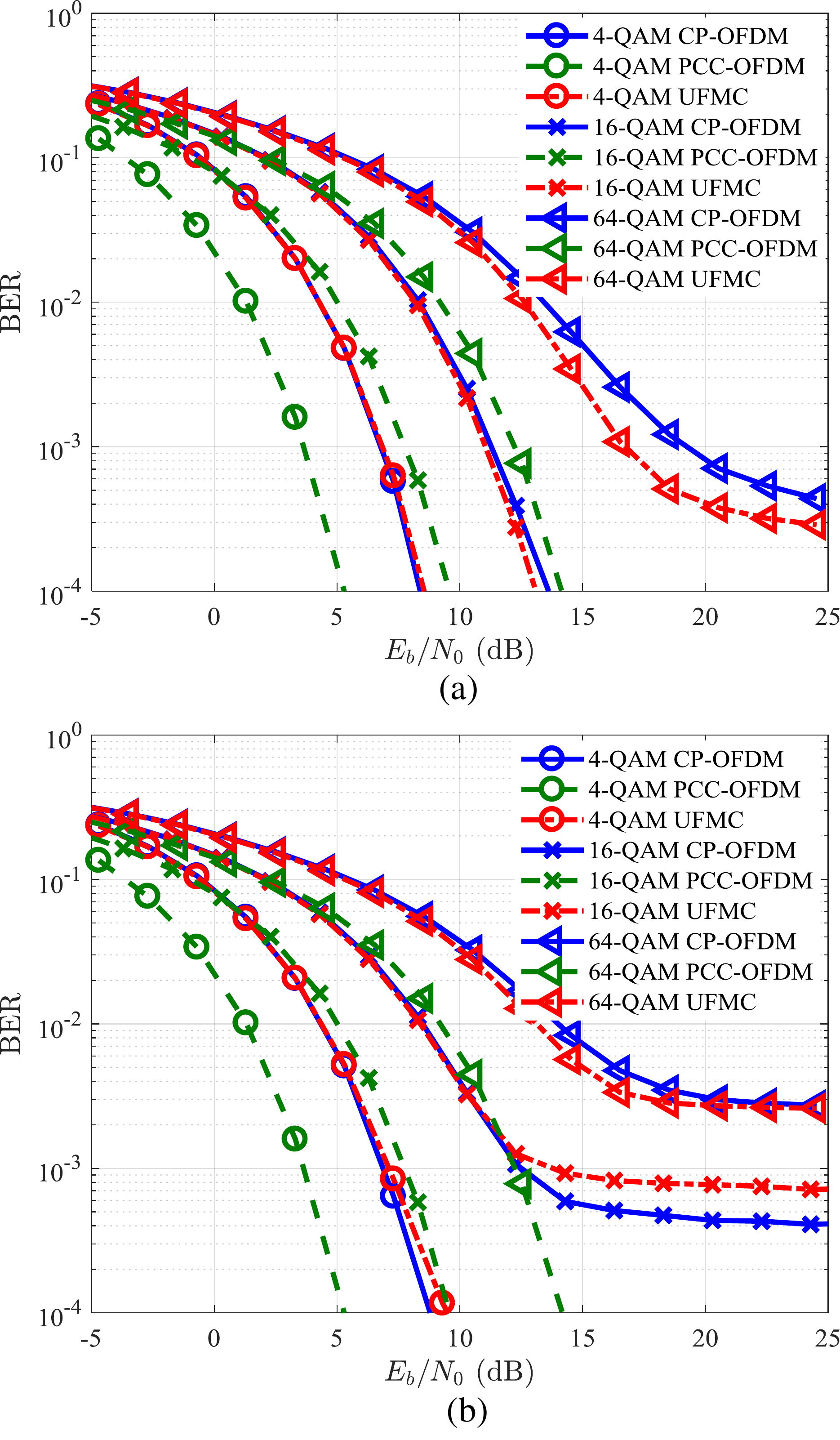}
\caption{Two user BER results for normalized frequency offset of User 2, $\Delta fT=0.2$, and normalize time offset of User 2, $\tau =0.05$. (a) guard band of 12 subcarriers, (b) no guard band.}
\label{fig:figure24}
\end{figure}

\section{Advantages and Disadvantages of PCC-OFDM for 5G}
\label{sec:advantage}
In addition to the advantages of extreme insensitivity to time and frequency described above, because PCC-OFDM is a form of OFDM, it also retains many of the advantages of CP-OFDM. For example, single tap equalizers can be used in PCC-OFDM to correct for frequency selective fading, and PCC-OFDM, like CP-OFDM is well suited to use in MIMO systems.  Although not considered in the simulations in this paper, PCC-OFDM has also been shown to be very insensitive to Doppler spread \cite{armstrong1998polynomial}.  The additional mapping onto subcarriers at the transmitter and the weighting-and-adding step at the receiver mean that PCC-OFDM adds negligible complexity and latency relative to CP-OFDM and has significantly lower complexity and latency than UFMC.

The most important potential disadvantage of the form of PCC-OFDM described in this paper is the potential loss in spectral efficiency resulting from the mapping of data onto pairs of subcarriers rather than onto single subcarriers. It is likely that much of this loss can be recovered due to a combination of the other properties of PCC-OFDM, including the fact that no frequency guard band is required between users, and that the improved performance may allow larger constellations to be used.  For low data rate applications a small loss in spectral efficiency may be acceptable.  For high data rate applications, particularly in the downlink, PCC-OFDM with overlapped symbol periods may offer a good solution \cite{armstrong2000performance}.

\section{Conclusions}
\label{sec:conclusion}
This paper has evaluated PCC-OFDM as a contender for 5G systems and compared it with well-known waveforms such as CP-OFDM and UFMC. It has been shown that PCC-OFDM performs substantially better than CP-OFDM and UFMC in the presence of time and/or frequency offsets. Large offsets that would make transmission with CP-OFDM or UFMC impossible cause very little degradation in PCC-OFDM. For example time offsets of $\pm 0.2$ of a symbol period require only a 2dB increase in SNR to achieve a given target BER.  Similarly frequency offsets of $\pm 0.2$of a subcarrier spacing also require an increase of only 2dB. It has been shown that the weighting-and-adding step in the PCC-OFDM receiver contributes significantly to the performance. Because of the mapping of data onto two subcarriers, weighting and adding improves the performance in AWGN by 3dB but it improves the performance by much more than this in the presence of time and frequency offsets. The performance of PCC-OFDM in a multi-access system using OFDMA has been evaluated. It has been shown that because PCC-OFDM is very well-localized in both the time and frequency domain it is extremely robust in a multiuser system and as a result no guard band is required between users. Only for very large constellations, very large frequency offsets, and when the power of the interfering user was 10dB higher than that of the user for which the BER was being calculated was any degradation observable. In conclusion PCC-OFDM is a very strong contender for application in 5G as it substantially outperforms other better known waveforms.

\bibliographystyle{IEEEtran}
\bibliography{PCCReferences}

\end{document}